\documentclass[10pt]{article}
\usepackage[cp866]{inputenc}
\usepackage[russian]{babel}
\begin{document}

%\inputencoding{cp866}

%\newcommand{\N}{N\raise.7ex\hbox{\underline{$\circ $}}$\;$}
\begin{center}

{\bf
BELARUS NATIONAL ACADEMY OF SCIENCES
\\
B.I. STEPANOV's INSTITUTE OF PHYSICS
}
\end{center}

\begin{center}
{\bf
V.V. Kisel,
N.G. Tokarevskaya,
A.A. Bogush,
V.M. Red'kov\footnote{E-Mail: redkov@dragon.bas-net.by}
}
\end{center}

\begin{center}
{\bf SPIN  1 PARTICLE IN A 15-COMPONENT FORMALISM, \\ INTERACTION WITH
ELECTROMAGNETIC AND  GRAVITATIONAL FIELDS }
\end{center}

\begin{center}
{\bf Abstract}
\end{center}

A generalized  vector particle theory with the use of an extended set of Lorentz group irredicible
representations, including scalar, two 4-vectors, and antisymmetric 2-rang tensor, is investigated
both in tensor an in matrix approaches. Initial equations depend upon four complex parameters $\lambda_{i}$,
obeying two sup\-plemen\-ta\-ry conditions, so restriction of the model to the case of electrically
neutral vector particle  is not a trivial task. A special basis in the space of 15-component wave functions
is found where instead of four $\lambda_{i}$ only one real-valued  quantity $\sigma$,
a bilinear combination of $\lambda_{i}$, is presented.  This $\lambda$-parameter is interpreted as an additional electromagnetic
characteristic of a charged vector particle, polarizability. It is shown that in this basis $C$-operation
is reduced to the complex conjugation only, without any accompanying linear transformation.
The form of $C$-operation in the initial
basis is calculated too. Invariant bilinear form matrix in both bases are found and the Lagrange formulation
of the whole theory is given. Explicit expressions of the conserved current vector  and
of the energy-momentum tensor are established. In presence of external electromagnetic fields, two
supple\-men\-ta\-ry field components, scalar and 4-vector, give a non-trivial contribution to the Lagrangian and
conserved quantities. Restriction to a  massless vector particle is determined.

Extension of the whole theory to the case of Riemannian space-time is accomp\-li\-shed. Two methods of obtaining
corresponding generally covariant wave equations are elaborated: of tensor- and of tetrad-based ones. Their equivalence is proved. It is shown that in case of pure curved space-time
models without Cartan torsion no specific additional interaction terms because of non-flat geometry arise.
The conformal symmetry of a massless generally covariant equation is demonstrated explicitly.
A canonical tensor of energy-momentum $T_{\beta \alpha}$  is constructed, its conservation law happens to involves
the Riemann curvature tensor. Within the framework of known ambiguity of any energy-momentum tensor,
a new tensor $\bar{T}_{\beta \alpha}$ is suggested to be  used, which obeys a common conservation law.

\vspace{5mm}

36 pages, references  --- 80

\vspace{10mm}


\vspace{5mm}

{\bf PASC numbers: } 1130, 0230, 0365, 2110H.

\newpage

\begin{center}
{\bf Content}
\end{center}

\noindent
\vspace{5mm}
Introduction

\noindent
\vspace{5mm}
1. Generalization of the vector particle theory

\noindent
\vspace{5mm}
2.  Special transformations of the wave function

\noindent
\vspace{5mm}
3.  Invariant bilinear form

\noindent
\vspace{5mm}
4.   On $C$-conjugation

\noindent
\vspace{5mm}
5.  Equations in a Riemannian space-time, tensor-based approach

\noindent
\vspace{5mm}
6. Covariant tetrad-based  formalism

\noindent
\vspace{5mm}
7.   Bilinear invariants in a curved space-time

\noindent
\vspace{5mm}
8.  Massless equation and conformal invariance.

\noindent
\vspace{5mm}
9.  Canonical energy-momentum tensor

\noindent
\vspace{5mm}
Results

\subsection*{Introduction}

In order to have a comprehensive  theory of higher-spin fields, it is necessary to be able to describe
interactions. The most important and the best understood is, of course,  the electromagnetic one. But the gravitation
is very important as well, at least from theoretical viewpoint if not from practical. Generally, a
standpoint may be brought  forward that we should give much attention to those facts of physics in the
Minkowski space-time which allow for  extension to a generally covariant physics.
Else one point should be taken as of principal importance. Since  wave equations with subsidiary
conditions usually lead to consistency  difficulties when  minimally coupled  to an electromagnetic  field,
it seems best to avoid subsidiary  conditions from the start. Furthermore, it seems the best to start with
a first-order system, which is  linked up to a certain Lagrange function, and in which some physically
required restrictive conditions  are built in from the very beginning in accordance with the  Pauli-Fierz
approach [1-2].   The more so is in  the context of possible background of a non-Euclidean geometry,
in view of arising additional subtleties in a consistency problem.

It is well known that if one looks for a first-order  differential equation describing a mass $m$ spin $s$
field,  that is form-invariant  under Lorentz transformations,  derivable from a Lagrangian, then  one  does not
uniquely obtain the common Dirac or Duffin-Kemmer or some other equations.
A general theory of such first-order and Lagrangian-based equations has been treated at
 great length
Fierz and Pauli [1-2], Dirac [3], Bhabha [4], Harish-Chandra [5], Umezawa [6],
and by many others [7-12].

In fact, it was shown that almost
an infinite many  of such equations is possible. However, there has been a notable dearth of examples
of such theories that have been developed at a large extent comparable
 to the Dirac or Duffin-Kemmer examples.
But  such  particle models, being elaborated in full  detail, might shed  new light on the theory of general
arbitrary-spin wave equations. Much work in this direction has been done [13-39].
For instance, generalized
wave equations for  particles of spin 1/2 and 1 respectively and with arbitrary anomalous magnetic momentums
have been worked out [15,17-19,21,22,24], [28,30,34,37-39]. Else one extension, less known, has been done for boson particle:
a next simplest theory beyond the
Duffin-Kemmer case was worked out  in [20,25,26,35,36] and an explicit representation of the basic first-order equation
was given.

In this paper we examine the properties of this new 15-component  spin 1 theory  in the case of no
interaction as well as  in presence of external electromagnetic and gravitational fields. It turns out that
this theory  is very simply related to the Duffin-Kemmer equation (in recent years some interest
in the Duffin-Kemmer-Petiau formalism again can be noted -- see [40-45]). In the free-field case,  the two theory are
equivalent, and the more complicated 15-component model can be reduced to the ordinary 10-component one.
In the presence of external electromagnetic fields, one again obtains  a Duffin-Kemmer like theory with an
additional term corresponding to an electromagnetic polarizability. Thus, this extended theory is definitely
inequivalent to the Duffin-Kemmer theory in presence of the minimal electromagnetic coupling. This else one
time illustrates explicitly a point emphasized in the literature that although two field theories  may be
equivalent in absence of interaction, they may be completely inequivalent in  presence of interaction.

\subsection*{1. Generalization of the vector particle theory}

The full physical content of an extended theory of a  spin 1  particle will emerge later. As for now,
we just postulate certain original set of the Lorentz group
 representation along with a wave equation for the
generalized  fields [20,25,26].
Initial generalized equation for a massive spin 1 particle, being associated with the scheme of links

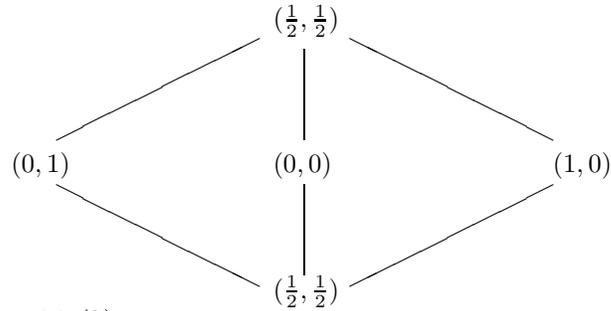
\unitlength= 0.60mm
\begin{picture}(0,+50)(-100,-10)
\special{em:linewidth 0.4pt}
\linethickness{0.4pt}

\put(-2,-2){$(0,0)$}
\put(-60,-2){$(0,1) $}
\put(+60,-2){$(1,0) $}
\put(-2,+30){$({1 \over 2},{1\over 2} ) $}
\put(-2,-30){$({1 \over 2},{1\over 2} ) $}

\put(-50,+5){\line(2,1){45}}
\put(-50,-5){\line(2,-1){45}}
\put(+60,+5){\line(-2,+1){45}}
\put(+60,-5){\line(-2,-1){45}}

\put(+5,+5){\line(0,+1){20}}
\put(+5,-5){\line(0,-1){20}}

\end{picture}

\vspace{12mm}

\noindent has the form\footnote{Latin indices  take the values $0,1,2,3$; at the work in the Minkowski
space-time the conventional  metric tensor $(g_{ab}) = diag(+1,-1,-1,-1); \;x_{0} = c t$ will be used.}
( $ m = Mc / \hbar$)
$$
\partial^{a}  (\lambda_{1} \Phi_{1a} + \lambda_{2} \Phi_{2a} )  - \Phi m  \ = 0 \; ,
$$
$$
\mp \lambda^{*}_{1} \partial_{a} \Phi + \lambda_{3} \partial^{b} \Phi_{ab} -  m \Phi_{1 a}  = 0 \; ,
$$
$$
\pm \lambda^{*}_{2} \partial_{a} \Phi + \lambda_{4} \partial^{b} \Phi_{ab} -  m   \Phi_{2 a} = 0 \; ,
$$
$$
\partial_{a} ( \lambda^{*}_{3} \Phi_{1 b} - \lambda^{*}_{4}  \Phi_{2 b} ) -
\partial_{b} ( \lambda^{*}_{3} \Phi_{1 a} - \lambda^{*}_{4}  \Phi_{2 a} ) -  m  \Phi_{ab}=0 \;.
\eqno (1.1)
$$

\noindent Here all the fields ($ \Phi , \Phi_{1a}, \Phi_{2a}, \Phi_{ab}$) as well as
$\lambda_{i}$ are complex-valued. Therefore restriction of the theory to the case of
chargeless particle, which cannot interact with external electromagnetic fields, is to be
a matter of special consideration.

It must be noted that the $\lambda_{i}$-s  are not fixed uniquely. Any set of such
$\lambda_{i}$ must obey two restricting conditions [25,26]
$$
\lambda_{1}\lambda^{*}_{1}-\lambda_{2}\lambda^{*}_{2}=0 \; , \qquad
\lambda_{3}\lambda^{*}_{3}-\lambda_{4}\lambda^{*}_{4}=1 \; .
\eqno (1.2)
$$

\noindent
In accordance with eqs. (1.2), freedom of choice of the $\lambda_{i}$-s is as follows:
$$
\left ( \begin{array}{c}
                        \lambda'_{1} \\ \lambda'_{2}
                                                      \end{array} \right  ) =
\left ( \begin{array}{cc}
                        \cosh \theta    & \sinh \theta  \\
                        \sinh \theta    & \cosh \theta
                                                       \end{array} \right )
\left ( \begin{array}{c}
                        \lambda_{1} \\ \lambda_{2}
                                                       \end{array} \right  ) ,
\left ( \begin{array}{c}
                         \lambda'_{3} \\ \lambda'_{4}
                                                       \end{array} \right ) =
\left ( \begin{array}{cc}
                        \cosh \Theta    & \sinh \Theta  \\
                        \sinh \Theta    & \cosh \Theta
                                                          \end{array} \right )
\left ( \begin{array}{c}
                        \lambda_{3} \\ \lambda_{4}
                                                     \end{array} \right )  \;    .
\eqno(1.3)
$$

\noindent
Generally speaking, $\theta-$ and $\Theta-$transformations are independent of each other.

The tensor equations  (1.1) can be formulated  as one matrix equation. To this end, let us  rewrite
eqs. (1.1) in the form
$$
\partial_{a} (\lambda _{1} g^{ab} ) \Phi_{1b} +
\partial_{a} ( \lambda_{2} g^{ab}) \Phi_{2b} -  m \Phi = 0  \; ,
$$
$$
\partial_{a} (\mp \lambda^{*}_{1} \delta^{a}_{k} ) \Phi -
\partial_{a} (\lambda_{3} g^{al} {1 \over 2} \;\delta_{lk}^{mn} ) \Phi _{mn} -  m   \Phi_{1k} = 0 \; ,
$$
$$
\partial_{a} (\pm \lambda^{*}_{2} \delta^{a}_{k} ) \Phi
-  \partial_{a} (\lambda_{4} g^{al} \delta_{lk}^{mn} ) \Phi _{mn} -  m \Phi_{2k} =0 \; ,
$$
$$
\partial_{a} ( \lambda^{*}_{3} \delta^{ab}_{kl} ) \Phi_{1b} +
\partial_{a} (-\lambda^{4}_{4} \delta_{kl}^{ab}) \Phi_{2b} -  m \Phi_{kl} = 0         \; .
\eqno(1.4)
$$

\noindent In eqs. (1.4) and in the following a special Kronecker's symbol $\delta^{ab}_{kl}$,
acting in the space of antisymmetric tensors as a unite operator
$$
\delta^{ab}_{kl} =  (\delta^{a}_{k}  \delta^{b}_{l}  - \delta^{a}_{l}  \delta^{b}_{k} ) \; , \;
{1 \over 2} \; \delta^{ab}_{kl} \; \Phi_{ab} = \Phi_{kl} \;.
$$

\noindent In turn, the equations (1.4) can be written as a matrix equation:
$$
\partial_{a}
\left ( \begin{array}{cccc}
0            &        \lambda_{1} g^{ab}    & \lambda_{2}g^{ab}        &           0         \\
\mp\lambda_{1}^{*} \delta^{a}_{k}      &    0     &     0   & -\lambda_{3} g^{al}\delta^{mn}_{lk} \\
\pm \lambda_{2}^{*} \delta^{a}_{k}   &   0  &     0    & -\lambda_{4} g^{al}\delta^{mn}_{lk} \\
0   &  +\lambda^{*}_{3} \delta^{ab}_{kl}  &  -\lambda^{*}_{4} \delta^{ab}_{kl}   &    0
\end{array} \right )
\left ( \begin{array}{l}
                     \Phi \\ \Phi_{1b} \\ \Phi_{2b} \\\Phi_{mn}
                                                           \end{array} \right ) =  m
\left ( \begin{array}{l}
                  \Phi \\ \Phi_{1k} \\ \Phi_{2k} \\\Phi_{kl}
                                                           \end{array} \right ) \; ,
\eqno(1.5a)
$$

\noindent or in a compact form
$$
(\; \Gamma^{a} \partial_{a}   -  m \; ) \; \Psi = 0 \; ,
\eqno(1.5b)
$$

\noindent where four matrices $\Gamma^{a}$ with the block structure
$$
\gamma^{a} = \left ( \begin{array}{cccc}
1             &    1 \times 4   &    1 \times 4     &   1  \times 6  \\
4 \times 1    &    4 \times 4   &    4 \times 4     &   4  \times 6  \\
4 \times 1    &    4 \times 4   &    4 \times 4     &   4  \times 6   \\
6 \times 1    &    6 \times 4   &    6 \times 4     &   6  \times 6
\end{array} \right )
\eqno(1.5c)
$$

\noindent have been used.
Take notice on the absence of a factor $1/2$ in second  row of the $\Gamma^{a}$ at $\lambda_{3}$;
this is so because
we have supposed that $\Psi$-function contains only 6 components of the antisymmetric tensor
$\Phi_{kl}$ (those are $ (01, 02, 03, 23, 31, 12$);  so one can  sum over 6-combinations $[kl]$ or can sum
over indices $k$ and $l$ independently but simultaneously inserting the factor $1/2$ (see (1.4)).

The freedom of choice (1.3) can be understood in the following way:
let us simultaneously with  the change $\lambda_{i} \Longrightarrow \lambda'_{i}$ (1.3) perform
non-degenerate linear transformation upon  $\Psi$-function:
$$
\Psi ' = S \; \Psi  ,  \;
S = S (\theta, \Theta ) \;   , \;\;
$$
$$
( \Gamma^{'a} \partial_{a} -  m  ) \; \Psi' = 0 ,   \;
\Gamma^{'a} = S  \Gamma^{a}  S^{-1}\; ,
\eqno(1.6a)
$$

\noindent Solving the equation $\Gamma^{'a} = S  \Gamma^{a}  S^{-1}\;$, where $\Gamma^{a}$ and $\Gamma^{'a}$
are functions of $\lambda_{i}$ and $\lambda_{i}'$ respectively (see (1.3)),  one can easily establish
the explicit form of $S$:
$$
\left ( \begin{array}{l}
\Phi'_{1а} \\ \Phi'_{2а}
\end{array} \right  ) =
\left ( \begin{array}{rr}
\cosh \theta    & -\sinh \theta  \\
-\sinh \theta    & \cosh \theta
\end{array} \right )
\left ( \begin{array}{l}
\Phi_{1а} \\ \Phi_{2а}
\end{array} \right ) \; , \;  \Theta = - \theta   \; .
\eqno(1.6b)
$$

Now we are coming to another question. Let us perform a special linear transformation
$$
\left ( \begin{array}{c}
                         C _{a}  \\ \Phi_{a}
                                                   \end{array} \right ) =
\left ( \begin{array}{rr}
                        \lambda_{1}  & \lambda_{2} \\
                        \lambda_{3}^{*} & -  \lambda_{4}^{*}
                                                   \end{array} \right )
\left ( \begin{array}{c}
                        \Phi_{1a}   \\ \Phi_{2a}
                                                   \end{array} \right ) \; , \;
$$
$$
\left ( \begin{array}{c}
                     \Phi_{1a}   \\ \Phi_{2a}
                                                   \end{array} \right ) =
{1 \over \lambda_{1} \lambda_{4}^{*} + \lambda_{2} \lambda_{3}^{*} }
\left ( \begin{array}{rr}
                     \lambda_{4}^{*}  & \lambda_{2} \\
                     \lambda_{3}^{*} & -  \lambda_{1}
                                                    \end{array} \right )
\left ( \begin{array}{c}
                    C _{a}  \\ \Phi_{a}
                                                    \end{array} \right ) \; .
\eqno(1.7)
$$

\noindent In these variables , eqs. (1.1) will take on the simpler form
(in the following will be convenient instead the scalar function $\Phi (x)$ to use
the notation  $C (x)$; so two supplementary components will be  $C(x)$ and $C_{a}$)
$$
\partial^{a} C_{a} -  m   C = 0  \;  , $$
$$
d \; \partial^{a} \Phi _{ba}  -  m  C_{b} = 0 ,
$$
$$
\mp d^{*} \; \partial_{b} C \; + \;\partial^{a} \Phi_{ba} -  m  \Phi_{b} = 0 \; ,
$$
$$
\partial_{a} \Phi_{b} - \partial_{b} \Phi_{a}  -  m   \Phi _{ab} = 0 \; ,
\eqno(1.8)
$$

\noindent where symbol $d$ stands for the combination of $\lambda_{i}$-s:
$ d = \lambda_{1} \lambda_{3} + \lambda_{2} \lambda_{4} \; $ .
Take notice that the new system (1.8) depends on $\lambda_{i}$ only through $d$ and  $d^{*}$.

In addition it is readily verified that  the introduced vectors $C(x)$ and $C_{a}(x)$ are
$\theta$-invariant. In other words, irrespective of a concrete choice of an initial  basis, fixed by
a set of $\lambda_{i}$, the final system (1.8) will be the same.

Eqs. (1.8) can be written, as well as eqs. (1.1), in a matrix form:
$$
(\Gamma^{a} \partial_{a}   -  m ) \Psi = 0 \;,
$$
$$
\partial_{a}
\left ( \begin{array}{cccc}
                 0         &  g^{ak}  &    0   &        0                       \\
                 0         &  0       &    0   &    - d g^{ab} \delta^{mn}_{bl} \\
\mp d^{*} \delta^{a}_{l}   &  0       &    0   &    -g^{ab} \delta^{mn}_{bl}    \\
                 0         &  0       &  \delta^{ak}_{bl}   &         0
                                                                               \end{array} \right )
\left ( \begin{array}{l}
                      C  \\ C_{k} \\ \Phi_{k}  \\ \Phi_{mn}
                                                                               \end{array} \right )
=  m  \left ( \begin{array}{l}
                          C  \\ C_{l} \\ \Phi_{l}  \\ \Phi_{bl}
                                                                               \end{array} \right ) \; .
\eqno(1.9)
$$

Having used two first equations in (1.8) one  can exclude the  supplementary field $C(x)$ from
third equation in (1.8):
$$
\mp         \; { d^{*} d  \over m^{2}  } \;  \partial_{a}\;
(\partial ^{k} \partial^{l} \Phi_{kl}) \; + \;\partial^{b} \Phi_{ab} -  m   \Phi_{a} = 0 \; ,
$$
$$
\partial_{a} \Phi_{b} - \partial_{b} \Phi_{a} -  m  \Phi _{ab} = 0\;.
\eqno(1.10a)
$$

\noindent As the term $(\partial ^{k} \partial^{l} \Phi_{kl})$ vanishes identically.
Thus, in free case (absence of external fields)
additional parameter of the extended theory of the spin 1 particle does not manifest itself anyhow.
At this two supplementary field components are as follows:
$$
C  = 0   \; , \; \partial^{a} C_{a} = 0 \; , \; C_{a} = d \; \Phi_{a} \; ,
\eqno(1.10b)
$$

\noindent and the main equations are
$$
\partial^{a} \Phi_{ba} -  m   \Phi _{b} = 0 \; , \;
\partial_{a} \Phi_{b} - \partial_{b} \Phi_{a} -  m   \Phi _{ab} = 0\; .
\eqno(1.11)
$$

\noindent these are ordinary Proca equations.

Other situation is realized in presence of external electromagnetic fields.
Making conventional change in eqs. (1.8)
$$
\partial _{a} \Longrightarrow  D_{a} =  \partial_{a}\; - \; i g \;  A_{a}(x)  \; ,
$$

\noindent $g = {e \over \hbar c},\; e$ --- electric charge of the particle, instead of (1.8) we get
$$
D^{a} C_{a} - m  C = 0  \; , \;       d \; D^{a} \Phi _{ ba} -  m  C_{b} = 0 \; ,
$$
$$
\mp d^{*} \; D_{b} C +  D^{a} \Phi_{ba} - m   \Phi_{b} = 0 \; , \;
$$
$$
D_{a} \Phi_{b} - D_{b} \Phi_{a} - m \Phi _{ab} = 0\;.
\eqno(1.12)
$$

\noindent From second equation in (1.12)  it follows
$$
D^{a}  C_{a}  =     { d  \over m}  D ^{a} D^{b} \Phi_{ab } =
i \;   {ed  \over m  } \; {1 \over 2} \;  F^{ ab } \; \Phi_{ab} \; ,
\eqno(1.13a)
$$

\noindent  where   $F_{ab} = ( \partial_{a} A_{b} - \partial_{b} A_{a})$.
Accounting for (1.13a), from first equation in  (1.12) we get
$$
C = - i \;  { e d \over m^{2} } \;{1 \over 2} \;  F^{ab} \Phi_{ab} \; .
\eqno(1.13b)
$$

\noindent
Substituting (1.13b) into third equation in  (1.12), we arrive    at
$$
D^{a} C_{a} - m C = 0  \; , \; \; d \; D^{a} \Phi _{ ba} - m  C_{b} = 0 \; ,
$$
$$
\pm i\; { e d d^{*}  \over  m^{2} } \; D_{k} \; ( {1 \over 2} \; F^{ab} \Phi_{ab} ) +
D^{a} \Phi_{a k}  - m \Phi_{k} = 0 \; ,
$$
$$
D_{a} \Phi_{b} - D_{b} \Phi_{a} - m  \Phi _{ab} = 0 \; .
\eqno(1.14)
$$

Equations (1.14) describe behavior of the vector particle in external electromagnetic field.
Two first relations in (1.14) are to considered as additional, they give us possibility
to construct supplementary fields $C(x), C_{a}$ in terms of the main $\Phi_a", \Phi_{ab}$.
In (1.14) $d^{*}d$-dependent term is an additional one and it represents some electromagnetic characteristic
of the particle.

Now we are coming to another question and will consider how one should determine the  massless  limit
of the theory. The corresponding equations are to be
$$
\partial^{a} C_{a} -   C  = 0  \; , \;
d \; \partial^{a} \Phi _{ ba } - C_{b} = 0 ,
$$
$$
\mp d^{*} \; \partial_{b} C +
\partial^{a} \Phi_{ba} = 0 \; , \;
$$
$$
\partial_{a} \Phi_{b} - \partial_{b} \Phi_{a} -
\Phi _{ab} = 0\; .
\eqno(1.15)
$$

\noindent This system is invariant under gauge transformation
$$
\Phi_{a} (x) \Longrightarrow \Phi '_{a}(x) =
\Phi _{a} (x) + \partial_{a} Z (x) \; ,
$$
$$
C(x) = inv \; , \;C_{a}(x) = inv \; , \;   \Phi_{ab}(x) = inv \; .
\eqno(1.16)
$$

\noindent   Having all the fields complex, the  $Z(x)$ is to be complex too. It should be borne
in mind that such a different behavior of two vector components $C_{a}, \Phi_{a}$ under gauge transformation
(see. (1.16) ) holds only in  the chosen basis. Taking the gauge transformation (1.16) to the initial basis
$\Phi_{1a}, \Phi_{2a}$ (1.1) we will have
$$
\Phi_{1a} \Longrightarrow \Phi_{1a} + {\lambda^{*}_{1} \over d^{*}} \; \partial_{a} \Lambda (x) \; , \;\;
\Phi_{2a} \Longrightarrow \Phi_{2a}  - {\lambda_{2}^{*} \over d^{*}} \; \partial_{a} \Lambda (x) \; ,
\eqno(1.17)
$$

\noindent both vectors transform simultaneously.

In the massless case the supplementary fields turn out to be zero
$$
C = 0  \; , \; C_{b} = 0 ,
$$

\noindent Thus, eq. (1.15) takes on the form of the ordinary Proca massless system
$$
\partial^{a} \Phi_{ba} = 0   \;  ,  \;  \partial_{a} \Phi_{b} - \partial_{b} \Phi_{a} - \Phi _{ab} = 0\; .
\eqno(1.18)
$$

\noindent This is quite understandable result, having remembered that in massive case the $\lambda_{i}$
exhibit themselves physically only in presence of external electromagnetic fields $A_{a}(x)$.

However, from heuristic considerations (having in mind the gauge invariance principle),
 in the massless case as well one can theoretical possibility,
 to study a massless complex-valued field in external vector
('electromagnetic') field\footnote{The term 'electromagnetic'  here should  be understood  with caution,
in fact as a matter of convention.}
At this eqs. (1.15) will change into
$$
D^{a} C_{a} -   C = 0 ,  \;
$$
$$
d \; D^{a} \Phi _{ ba} - C_{b} = 0 \; , \;
$$
$$
\mp d^{*} \; D_{b} C  + D^{a} \Phi_{ba} = 0 \; ,  \;
$$
$$
D_{a} \Phi_{b} - D_{b} \Phi_{a} - \Phi _{ab} = 0\; ,
\eqno(1.19)
$$

\noindent where $D_{a} = \partial _{a} - i g  A_{a}$.
Eqs. (1.19) are invariant under $U(1)$-gauge transformation, different from previous gauge symmetry (1.16):
$$
A_{a} \Longrightarrow A_{a} + \partial_{a} \Lambda (x) \; , \;
(C, C_{a}, \Phi_{a}, \Phi_{ab}) \Longrightarrow
e^{ig \Lambda } \; (C, C_{a}, \Phi_{a}, \Phi_{ab}) \; .
\eqno(1.20)
$$

It can be readily verified that eqs. (1.19) are equivalent to the following ones
$$
C_{b} = d \; D^{a} \Phi_{ba} \; ,
$$
$$
C =   - i \;  e d \; {1 \over 2} \; F^{ab} \Phi_{ab} \; ,
$$
$$
\pm i \; e d^{*} d \;   D_{b} ( {1 \over 2 }\; F^{kl}\Phi_{kl}) + D^{a} \Phi_{ab} = 0 \; ,
$$
$$
D_{a} \Phi_{b} - D_{b} \Phi_{a} - \Phi _{ab} = 0\; .
\eqno(1.21)
$$

So, in a massless case, in presence of external vector fields $A_{a}$, the $\lambda_{i}$-parameters
manifest themselves as well.

\subsection*{2. Special transformations of the wave function}

In  Sec. 2  we are going to give special attention to the question of the charge-symmetry.
At this, for simplicity,  instead of $D_{a} = \partial_{a} -i g A_{a}$) we will write  the operator
$\partial_{a}$, but  remembering  that everything concerns the charged particle as well
(with taking $g$ into $-g$ at $C$-conjugation).

Let us turn to eqs. (1.8) and write down complex-conjugate ones:
$$
\partial^{a} C^{*}_{a} - m  C^{*} = 0  \; , \;
$$
$$
d^{*} \; \partial^{a} \Phi^{*} _{ ba}  -  m  C^{*}_{b} = 0 ,
$$
$$
\mp d \; \partial_{b} C^{*} + \partial^{a} \Phi^{*}_{ba} -  m  \Phi^{*}_{b} = 0 \; , \;
$$
$$
\partial_{a}  \Phi^{*}_{b}  -  \partial_{b}  \Phi^{*}_{a}  -   m  \Phi^{*} _{ab} = 0  \;  .
\eqno(2.1)
$$

\noindent
Defining  $C$-conjugate wave functions according to
$$
(d / d^{*} ) C^{*} = C^{(c)} \; , \;  (d / d ^{*}) C^{*}_{a} = C^{(c)}_{a} \; , \;
$$
$$
\Phi_{a}^{*} = \Phi^{(c)}_{a} \; , \;  \Phi_{ab}^{*} = \Phi^{(c)}_{ab} \; , \;
\eqno(2.2)
$$

\noindent eqs.  (2.1)  can be rewritten as
$$
\partial^{a} C^{(c)}_{a} -  m  C^{(c)} = 0  \; , \;
d \; \partial^{a} \Phi^{(c)} _{ba}  - m   C^{(c)}_{b} = 0 ,
$$
$$
\mp d^{*} \; \partial_{b} C^{(c)} +  \partial^{a} \Phi^{(c)}_{ba} - m  \Phi^{(c)}_{b} = 0 \; , \;
$$
$$
\partial_{a} \Phi^{(c)}_{b}  - \partial_{b} \Phi^{(c)}_{a}  - m  \Phi^{(c)} _{ab} = 0 \; ,
$$

\noindent which exactly coincides with eqs. (1.8).  Therefore eqs. (1.8) are $C$-invariant and
$C$-conjugation matrix has the form
$$
\Psi^{c} = C \; \Psi^{*}  \; , \;
C = \left ( \begin{array}{cccc}
                       (d / d^{*} ) & 0 & 0 & 0 \\
                       0  & (d / d^{*}) I_{4} & 0 & 0 \\
                       0  &  0  &  I_{4}  & 0   \\
                       0  &  0  &  0  &  I_{6}
                                                           \end{array} \right ) \; .
\eqno(2.3)
$$

Now, let us show  that there exists  a basis in which $S$-operation has a simpler form.  To this end,
turning again to eqs. (1.8), let us  separate out phase factors from  $d$  and $d^{*}$:
$ d \Longrightarrow  d_{0} e^{i t} \; , \; d^{*} \Longrightarrow  d_{0} e^{-i t} \; . $
Then eqs. (1.8) will read as
$$
\partial^{a} C_{a} - m   C = 0  \; , \;
$$
$$
d_{0} e^{+i t} \; \partial^{a} \Phi _{ba}  - m C_{b} = 0  \; ,
$$
$$
\mp d_{0} e^{-it} \; \partial_{b} C + \partial^{a} \Phi_{ba} - m \Phi_{b} = 0 \; , \;
$$
$$
\partial_{a} \Phi_{b} - \partial_{b} \Phi_{a}  - m \Phi _{ab} = 0 \; .
\eqno(2.4)
$$

\noindent Now, defining new fields
$$
e^{-it} C (x) \Longrightarrow C (x) \; , \; e^{-it} C_{a} (x) \Longrightarrow  C_{a} (x) \; , \;
\eqno(2.5)
$$

\noindent we take eqs.  (2.4) into
$$
\partial^{a} C_{a} - m C = 0  \; , \;
$$
$$
d_{0} \; \partial^{a} \Phi _{ba}  - m    C_{b} = 0 \;  ,
$$
$$
\mp d_{0} \; \partial_{b} C +
\partial^{a} \Phi_{ba} - m  \Phi_{b} = 0 \; , \;
$$
$$
\partial_{a} \Phi_{b} - \partial_{b} \Phi_{a}  - m \Phi _{ab} = 0 \; ,
\eqno(2.6)
$$

\noindent here $d_{0}$ is a real-valued quantity.

Matrix form of eqs. (2.6) is
$$
\partial_{a}
\left ( \begin{array}{cccc}
                  0    &    g^{ak}  &     0     &                   0                       \\
                  0    &    0       &     0     &   - d_{0} g^{ab} \delta^{mn}_{bl}         \\
\mp d_{0} \delta^{a}_{l} &  0       &     0     &   - g^{ab} \delta^{mn}_{bl}               \\
                  0    &    0       &   \delta^{ak}_{bl}   &        0
  \end{array} \right )
\left ( \begin{array}{l}
                         C \\ C_{k} \\ \Phi_{k}  \\ \Phi_{mn}
\end{array} \right )
= m  \left ( \begin{array}{l}
                             C  \\ C_{l} \\ \Phi_{l}  \\ \Phi_{bl}
\end{array} \right ) \; .
\eqno(2.7)
$$

\noindent Evidently, in this basis  $C$-operation is reduced to the complex conjugation only:
$$
\Psi^{c}(x) =  \; \Psi^{*} (x) \; .
\eqno(2.8)
$$

Else one transformation upon $\Psi $ is to be done:
$$
d_{0}^{-1} \; C (x)   \Longrightarrow C' (x) \; , \;
d_{0}^{-1} \; C_{a} (x)   \Longrightarrow C'_{a} (x) \; ,
\eqno(2.9)
$$

\noindent then eqs. (2.6) read as
$$
\partial^{a} C_{a} - m  C = 0  \; , \;\;
$$
$$
\partial^{a} \Phi _{b a }  - m  C_{b} = 0 ,
$$
$$
\mp d_{0}^{2} \; \partial_{b} C +
\partial^{a} \Phi_{ba} - m   \Phi_{b} = 0 \; , \;
$$
$$
\partial_{a} \Phi_{b} - \partial_{b} \Phi_{a}  - m  \Phi _{ab} = 0 \; .
\eqno(2.10)
$$

In the following this $ d_{0}^{2}$ will be designated as $ \sigma $. Also it is quite understandable
that  existence of two possibilities in eqs. (2.10), associated with two signs $\pm$, can be described
as different in sign quantities  $ \sigma$. In other words, these signs $\pm$ at $\sigma$  can be omitted.
It should be mentioned that  $\sigma$-parameter is dimensionless whereas all the component of $\Psi$ have
dimension ($l = cm $:
$
[ C ] = [ C_{a} ] = [ \Phi_{a} ] = [ \Phi_{ab} ] = l^{-3/2}  \; $.

\noindent
Matrix form of eqs. (2.10)  is as follows
$$
\partial_{a}
\left ( \begin{array}{cccc}
                        0     &     g^{ak}  &     0   &                0                \\
                        0     &       0     &     0   &    -g^{ab} \delta^{mn}_{bl}     \\
        \sigma \delta^{a}_{l} &       0     &     0   &    - g^{ab} \delta^{mn}_{bl}    \\
                        0     &       0     &  \delta^{ak}_{bl}    &   0
\end{array} \right )
\left ( \begin{array}{l}
C \\ C_{k} \\ \Phi_{k}  \\ \Phi_{mn}
\end{array} \right ) =  m
\left ( \begin{array}{l}
                   C \\ C_{l} \\ \Phi_{l}  \\ \Phi_{bl}
\end{array} \right ) .
\eqno(2.12)
$$

\noindent In massless limit, instead of (2.11) we will have
$$
\partial^{a} C_{a} -   C = 0  \; , \;
$$
$$
\partial^{a} \Phi _{b a }  -   C_{b} = 0 \; ,
$$
$$
\sigma \; \partial_{b} C + \partial^{a} \Phi_{b a}  = 0 \; , \;
$$
$$
\partial_{a} \Phi_{b} - \partial_{b} \Phi_{a}  -   \Phi _{ab} = 0 \; .
\eqno(2.13)
$$

To avoid misunderstanding  it should be noted that thought the same symbol $\sigma$
is used both in (2.11) and (2.13)
but this symbol stands for quite different characteristics. In massless case $\sigma$ must have
dimension of squared length.

\subsection*{3.  Invariant bilinear form }

Let us consider the question about explicit expression of invariant bilinear form matrix in this theory.
In the work with the equation (see (2.12))
$$
(\Gamma^{a} \partial_{a}   - m  )\; \Psi = 0 \;
\eqno(3.1)
$$

\noindent we will employ the block structure of the $\Gamma^{a}$-s
$$
\Gamma^{a} =
\left ( \begin{array}{llll}
                 0          &  G^{a}     &  0     &  0  \\
                 0          &  0         &  0     &  K^{a} \\
         \sigma \Delta^{a}  &  0         &  0     &  K^{a} \\
                 0          &  0         &  \Lambda ^{a}  &  0
\end{array} \right ) \; ,
\eqno(3.2a)
$$

\noindent where
$$
(K^{a}) _{l} ^{\;\;mn} =
(- g^{am} \delta^{n}_{l} + g^{an} \delta^{m}_{n}) \; , \;
$$
$$
(\Lambda^{a})_{bl}^{\;\;\;\;k} = \delta_{bl}^{ak} \; , \;
$$
$$
(G^{a})_{(0)}^{\;\;\;k} = g^{ab} \; , \;\;
$$
$$
(\Delta^{a})^{\;\;\;(0)}_{l} =  \delta^{a}_{l} \; .
\eqno(3.2b)
$$

As known, required matrix $\eta $  of invariant form is determined by the relation  [12]
$$
\eta^{-1} (\Gamma^{a})^{+} \eta  = - \Gamma^{a} \; .
\eqno(3.3)
$$

\noindent If the $\eta$  is found, then $\bar{\Psi}  = \Psi^{+} \eta $  will obey
$$
\bar{\Psi } \; (\Gamma^{a} \stackrel{\leftarrow}{\partial}_{a} + m )  = 0 \; .
\eqno(3.4)
$$

\noindent Allowing  for (3.1) and (3.4), one can straightforwardly obtain this conventional expression
for a current conserved:
$$
J^{a} = \bar{\Psi} \Gamma^{a} \Psi \; , \; \partial_{a}  J^{a} = 0 \; .
\eqno(3.4)
$$

By definition, a Lorentz invariant is to be constructed with the use of $\eta$ according to
formula $\Psi ^{+} \eta \Psi$. Its most general structure is as follows
$$
\Psi ^{+} \eta \Psi =
r_{1} \; C^{*} C  +  r_{2} \; C^{*}_{a} C^{a}  +  r_{3} \; C^{*}_{a} \Phi^{a} +
r_{4} \;\Phi^{*}_{a} C^{a} + r_{5} \; \Phi^{*}_{a} \Phi^{a} +  {1 \over 2} r_{6} \;
 \Phi^{*} _{ab} \Phi _{ab}  \; ,
$$

\noindent and hence  $\eta$ can be as
$$
\eta  =
\left ( \begin{array}{llll}
                      R_{1}  &  0  &  0      &  0  \\
                      0  &  R_{2}  &  R_{3}  &  0  \\
                      0  &  R_{4}  &  R_{5}  &  0  \\
                      0  &  0      &  0      &  R_{6}
\end{array}  \right ) =
\left ( \begin{array}{cccc}
                      r_{1}  &  0  &  0      &  0  \\
                      0  &  r_{2} g^{ab}  &  r_{3} g^{ab} &  0  \\
                      0  &  r_{4} g^{ab}  &  r_{5} g^{ab} &  0  \\
                      0  &  0      &  0      &  r_{6} H^{mn,ab}
\end{array}  \right ) \; .
\eqno(3.5)
$$

\noindent Here  $H^{mn, ab} = (g^{ma} g^{nb} - g^{mb} g^{na}) \; $ .
Taking into account representation of  $\Gamma^{a}, (\Gamma^{a})^{+}$ and  $\eta$
in terms of block-matrices, from eq. (3.3) rewritten as
$ \Gamma^{a+} \eta  + \eta  \Gamma^{a} = 0 \; $, we readily get to the set of relations
$$
\sigma  \tilde{\Delta}^{a} R_{4}  +    R_{1}  G^{a} = 0 \; , \;\; R_{5} = 0 \; ,
$$
$$
\tilde{G}^{a} R_{1} + \sigma R_{3} \Delta^{a} \; , \;\; R_{2} + R_{3} = 0 \; , \;
$$
$$
R_{5} = 0  \; ,  \;\;  \tilde{\Lambda}^{a} R_{6}  +    R_{4}  K^{a}  = 0 \; ,
$$
$$
 R_{2} + R_{4} = 0 \; , \;\; \tilde{K}^{a} R_{3} +     R_{6} \Lambda ^{a} = 0 \; .
\eqno(3.6a)
$$

\noindent In (3.6) one can distinguish couples of equations transformed into each other
on matrix transposing;  therefore each couple provides us with a single equation.
Allowing for formulas (3.2), from eqs. (3.6) it follows
$$
r_{1} = + \; \sigma \; r_{2} \; , \; r_{3} = -  \; r_{2} \; ,
\; r_{4} = - \; r_{2} \; , \; r_{5} = 0 \; , \;r_{6} = - \; r_{2} \; .
\eqno(3.6b)
$$

\noindent Thus, in the used basis  $\eta$-matrix  is given by
$$
\eta =  \; const \;\left ( \begin{array}{llll}
                        +  \; \sigma  &  0  &  0      &  0  \\
                        0  &  +  g^{ab}  &  -   g^{ab} &  0  \\
                        0  &  -  g^{ab}  &  0  &  0  \\
                        0  &  0      &  0      &  -  H^{mn,ab}
\end{array}  \right ) \; .
\eqno(3.7)
$$

\noindent The relevant feature of $\eta$-matrix is that the invariant-form matrix has turned out to be
of non-diagonal structure.

But this property does not hold in all other possible bases. For instance, let us show that in
a representation associated with eqs. (1.1) the $\eta$-matrix will have a common diagonal form.
Actually, being linked up to eqs. (1.8), the $\Gamma^{a}$-s are
$$
\Gamma^{a} = \left ( \begin{array}{cccc}
           0     &   \lambda_{1} G^{a}   &  \lambda_{2}  G^{a}   &  0 \\
           - \nu \lambda_{1}^{*} \Delta^{a}  &  0  &  0  &\lambda_{3} K^{a} \\
           + \nu \lambda_{2}^{*} \Delta^{a}  &  0  &  0  &\lambda_{4} K^{a} \\
           0  &  \lambda_{3}^{*} \Lambda^{a}  & - \lambda_{4}^{*} \Lambda^{a} & 0
\end{array} \right )   \; ,
\eqno(3.8)
$$

\noindent here the same notation  for block-matrices $G^{a}, \Delta^{a}, K^{a},\Lambda^{a}$ as in
(3.2b) $\eta$ has been used. In contrast to the above, here an appropriate  substitution for $\eta$-matrix
happens to be of typical diagonal form
$$
\eta =
\left ( \begin{array}{llll}
                        C_{1}   &  0  &  0  &  0  \\
                        0  &  C_{2}  &  0  &  0   \\
                        0  &  0  &   C_{3} &  0 \\
                        0  &  0  &  0  & C_{4}
\end{array} \right ) =
\left ( \begin{array}{llll}
                        c_{1}   &  0  &  0  &  0  \\
                        0  &  c_{2} g^{ab}   &  0  &  0   \\
                        0  &  0  &   c_{3} g^{ab}  &  0 \\
                        0  &  0  &  0  & c_{4} H^{ab,mn}
\end{array} \right ) \; .
\eqno(3.9)
$$

\noindent Correspondingly, the relation  $(\Gamma^{a})^{+} \eta  + \eta  \Gamma^{a} = 0 \; $ yields
$$
-  \nu \tilde{\Delta}^{a}C_{2}  +   C_{1} G^{a}  = 0 \; , \;\;
\nu \tilde{\Delta}^{a}C_{3} +  C_{1} G^{a}   = 0 \; ,
$$
$$
\tilde{G}^{a}C_{1}  -  \nu C_{2} \Delta^{a} \; , \;\;
\tilde{\Lambda}^{a} C_{4}  +  C_{2} K^{a}  \; ,
$$
$$
\tilde{G}^{a} C_{1}  +  \nu C_{3} \Delta^{a}   = 0 \; , \;\;
- \tilde{\Lambda}^{a} C_{4} +  C_{3} K^{a}   = 0 \; ,
$$
$$
 \tilde{K}^{a}C_{2}  +  C_{4} \Lambda^{a}  = 0 \; , \;    \;
\tilde{K}^{a}C_{3}  -  C_{4} \Lambda^{a}    = 0 \; ,
$$

\noindent from which one can easily derive
$$
c_{2} = + c_{1} \nu \; , \;\; c_{3} = - c_{1} \nu  \; , \;\;  c_{4} = + c_{1}  \nu \; .
$$

\noindent Therefore, we have got to
$$
\eta = c  \;  \left (  \begin{array}{cccc}
                   - \nu  &  0  &  0  &  0 \\
                     0  & - g^{ab}  & 0  &  0  \\
                     0  &  0  &  + g^{ab}  & 0 \\
                     0 &  0  &  0 &  - H^{ab,mn}
\end{array} \right ) \; .
\eqno(3.10)
$$

One can give else one (additional) treatment of the matter: namely, to follow how the expression for $\eta$
in the final basis (3.7)  might be calculated on translating that of the initial $\eta$ according to (3.10).
The rule for transforming an invariant-form  matrix at  changing  bases in $\Psi$-space can be derived quite
easily. From relations defining $\eta$:
$$
\bar{\Psi} = \Psi^{+} \eta \; , \qquad \eta^{-1} \Gamma^{a+} \eta = - \Gamma^{a} \; ,
$$

\noindent will immediately  follow analogous ones for defining $\eta'$
$$
\bar{\Psi}' = \Psi^{'+} \eta' \; ,  \quad   (\eta ')^{-1} \Gamma^{'a+} \eta' = - \Gamma^{'a} \; ,
$$

\noindent if the following relationships
$$
\eta ' = (S^{+})^{-1} \eta \; S^{-1} \; , \;\;\;\; or  \;\;\;\;
\eta  = (S^{+}) \; \eta ' \; S  \; .
\eqno(3.11)
$$

\noindent hold.

With the use of eq. (3.11) let us transform the  $\eta $-matrix (3.10) to a new basis.
At this it will be convenient to take in mind the following scheme  (see.  Sec  2)

\vspace{5mm}

\unitlength= 0.60mm
\begin{picture}(0,+25)(-30,-10)
\special{em:linewidth 0.4pt}
\linethickness{0.4pt}

\put(0,0){$\lambda_{i}$}

\put(+15,0){$\stackrel{S_{1}}{\Longrightarrow}$}
\put(30,0){$\{d , d^{*}\} $}

\put(+55,0){$\stackrel{S_{2}}{\Longrightarrow}$}
\put(70,0){$\{\mid d \mid , \mid d \mid \} $}

\put(+110,0){$\stackrel{S_{3}}{\Longrightarrow}$}
\put(125,0){$ \{\sigma =  \mid d \mid ^{2} \} $}

\put(-3,-10){$\Psi , \eta $}
\put(33,-10){$\Psi', \eta' $}
\put(77,-10){$\Psi'', \eta '' $}
\put(130,-10){$\Psi''', \eta''' $}

\end{picture}

\vspace{5mm}

First,  let us perform the transformation $\Psi ' = S_{1} \Psi \;$. It will be more convenient to employ
the formula  $ S_{1}^{+} \eta' \; S_{1} = \eta \; $, where
$$
S_{1} = \left ( \begin{array}{ccrc}
1 & 0  &  0 & 0 \\
0  &  \lambda_{1}  &  \lambda_{2}  &  0  \\
0  &  \lambda_{3}^{*}  &   - \lambda_{4}^{*} &  0  \\
0  &  0  &  0  &  I_{6}   \end{array} \right ) \; , \;
S_{1}^{+}  = \left ( \begin{array}{ccrc}
1 & 0  &  0 & 0 \\
0  &  \lambda_{1}^{*}  &  \lambda_{3}  &  0  \\
0  &  \lambda_{2}^{*}  &   - \lambda_{4} &  0  \\
0  &  0  &  0  &  I_{6}   \end{array} \right ) \; .
\eqno(3.12)
$$

\noindent Thus,  allowing for the substitution for $\eta'$
$$
\eta ' =
\left ( \begin{array}{llll}
R_{1}  &  0  &  0      &  0  \\
0  &  R_{2}  &  R_{3}  &  0  \\
0  &  R_{4}  &  R_{5}  &  0  \\
0  &  0      &  0      &  R_{6}
\end{array}  \right ) \; ,
$$
\noindent  with the use of the relations $ S_{1}^{+} \eta' \; S_{1} = \eta $ and (see.  (3.10))
$\ r_{1} = - c \nu \; ,  R_{6} = c \; H   \; $, one can derive the following
$$
(\lambda_{1}^{*} R_{2} + \lambda_{3} R_{4} ) \; \lambda_{1} +
(\lambda_{1}^{*} R_{3} +\lambda_{3} R_{5}  ) \; \lambda_{3}^{*}   = - c \; ,
$$
$$
(\lambda_{1}^{*} R_{2} + \lambda_{3} R_{4} ) \; \lambda_{2} -
(\lambda_{1}^{*} R_{3} +\lambda_{3} R_{5}  ) \; \lambda_{4}^{*} = 0 \; ,
$$
$$
(\lambda_{2}^{*} R_{2} - \lambda_{4} R_{4} ) \; \lambda_{1} +
(\lambda_{2}^{*} R_{3} -\lambda_{4} R_{5}  ) \; \lambda_{3}^{*} = 0 \; ,
$$
$$
(\lambda_{2}^{*} R_{2} - \lambda_{4} R_{4} ) \; \lambda_{2} -
(\lambda_{2}^{*} R_{3} -\lambda_{4} R_{5}  ) \; \lambda_{4}^{*} = + c \; .
\eqno(3.13)
$$

\noindent
With the notation
$$
\left. \begin{array}{l}
\lambda_{1}^{*} r_{2} + \lambda_{3} r_{4}    = A  \\
\lambda_{2}^{*} r_{2} - \lambda_{4} r_{4}    = a
\end{array} \right.    \; , \;\;
\left.  \begin{array}{l}
\lambda_{1}^{*} r_{3} + \lambda_{3} r_{5}    = B   \\
\lambda_{2}^{*} r_{3} - \lambda_{4} r_{5}    = b
\end{array} \right.  \;
\eqno(3.14)
$$

\noindent the system (3.13) reads as
$$
\left \{ \begin{array}{l}
A \lambda_{1} + B \lambda_{3}^{*} = - c  \\
A \lambda_{2} - B \lambda_{4}^{*} = 0
\end{array} \right. \; , \;
\left \{ \begin{array}{l}  a \lambda_{1} + b \lambda_{3}^{*} =  0  \\
a \lambda_{2} -  b \lambda_{4}^{*} = + c
\end{array} \right.  \; .
\eqno(3.15)
$$

\noindent From (3.15)  it follows
$$
A = - { \lambda_{4}^{*} \over \lambda_{1} \lambda_{4}^{*} +
\lambda_{2} \lambda_{3}^{*} } \;  c \; , \;\;
B = - { \lambda_{2} \over \lambda_{1} \lambda_{4}^{*} +
\lambda_{2} \lambda_{3}^{*} } \; c \; , \;\;
$$
$$
a = + { \lambda_{3}^{*} \over \lambda_{1} \lambda_{4}^{*} +
\lambda_{2} \lambda_{3}^{*} } \; c \; , \;\;
b = - { \lambda_{1} \over
\lambda_{1} \lambda_{4}^{*} +
\lambda_{2} \lambda_{3}^{*} } \; c \; ,
$$

\noindent In turn, from  (3.14) we obtain
$$
r_{2} = { A \lambda_{4} + a \lambda_{3} \over
\lambda_{1}^{*}  \lambda_{4} + \lambda_{2}^{*} \lambda_{3} } \; , \;
r_{4} = { A \lambda_{2}^{*} - a \lambda_{1}^{*} \over
\lambda_{1}^{*}  \lambda_{4} + \lambda_{2}^{*} \lambda_{3} } \; , \;
$$
$$
r_{3} = { B \lambda_{4} + b \lambda_{3} \over
\lambda_{1}^{*}  \lambda_{4} + \lambda_{2}^{*} \lambda_{3} } \; , \;
r_{5} = { B \lambda_{2}^{*} - b \lambda_{1}^{*} \over
\lambda_{1}^{*}  \lambda_{4} + \lambda_{2}^{*} \lambda_{3} } \; .
$$

\noindent  From this, accounting for the above expressions for  $A,B,a,b$, we get to
$$
r_{2} =   { + c \over D D^{*} }  \; , \qquad
r_{4} =  { - c \over D D^{*} } \; d ^{*} \; , \qquad
r_{3} =  { - c \over D D^{*} } \;d \; , \qquad
r_{5} =  0 \; ,
\eqno(3.16)
$$

\noindent  where  $ D = \lambda_{1}^{*} \lambda_{4} +  \lambda_{2} ^{*} \lambda_{3} $.
With  $\lambda_{1} \lambda_{1}^{*} = \lambda_{2} \lambda_{2}^{*}$,
the expression for $D$  can be rewritten as
$$
D = \lambda_{1}^{*} \lambda_{4} + \lambda_{2} ^{*} \lambda_{3}  =
\lambda_{1}^{*} ( \lambda_{4} + { \lambda_{2}^{*} \over \lambda_{1}^{*} }\; \lambda_{3}  ) =
\lambda_{1}^{*}  ( \lambda_{4} + { \lambda_{1} \over \lambda_{2} } \;
\lambda_{3}) = { \lambda_{1} \over    \lambda_{2} } \; d \; ;
$$

\noindent and correspondingly the equality
$$ D \; D^{*} = { \lambda_{1} \over    \lambda_{2} } \; d \;
                { \lambda_{1}^{*} \over    \lambda_{2}^{*} } \; d^{*} =  d \; d^{*}  \;
$$

\noindent holds. Therefore, we have arrived  at the result:
$$
r_{1} = - c \nu  \; , \;  r_{6} = + c \; , \;
r_{2} = + {1 \over d^{*}d} \; c \; , \;
r_{3} = - {1 \over d^{*}} \; c \; , \;
r_{4} = - {1 \over d} \; c \; , \;
r_{5} = 0 \; ,
\eqno(3.17a)
$$

\noindent that is
$$
\eta ' = c \;
\left (  \begin{array}{cccc}
      -\nu    &            0              &         0                 &  0    \\
0             & (d d^{*})^{-1} g^{ab}     &    - (d^{*})^{-1} g^{ab}  &  0    \\
0             &  - d^{-1} g^{ab}          &         0                 &  0    \\
0             &            0              &         0                 & + H^{ab,mn}
\end{array} \right ) \; .
\eqno(3.17b)
$$

Now let us perform else one transformation
$$
\Psi '' = S_{2} \Psi  ' \; , \;\;
S_{2} = \left ( \begin{array}{cccc}
              e^{-it}  &  0              &      0     &  0  \\
                0      &  e^{-it} I_{4}  &      0     &  0  \\
                0      &  0              &    I_{4}   &  0  \\
                0      &  0              &      0     &  I_{6}
\end{array} \right ) \; .
\eqno(3.18a)
$$

\noindent Correspondingly for $\eta''$ we will have
$$
\eta'' =  c \;\left (  \begin{array}{cccc}
-\nu  &  0  &  0  &  0  \\
0  & d_{0}^{-2}  g^{ab}   &    - d_{0}^{-1}  g^{ab}  &  0    \\
0  &  - d_{0}^{-1}  g^{ab}  &   0  &  0  \\
0  &  0  &  0 & + H^{ab,mn}
\end{array} \right ) \; .
\eqno(3.18b)
$$

\noindent
And finally, third step:
$$
C''' = d_{0}^{-1} C '' \; , \; C_{a} ''' = d_{0}^{-1} \Phi_{a} C_{a}'' \; , \;
$$
$$
\Phi_{a}''' = \Phi_{a}''\; , \; \Phi_{ab}''' = \Phi_{ab}''\; ,
\eqno(3.19a)
$$

\noindent  and further
$$
\eta''' =
c \;\left (  \begin{array}{cccc}
                - \nu d_{0}^{2}         &          0         &         0         &          0  \\
                     0              &    + g^{ab}        &    -   g^{ab}     &          0  \\
                     0              &    -  g^{ab}       &         0         &          0  \\
                     0              &          0         &         0         &  - H^{ab,mn}
\end{array} \right )    \; .
\eqno(3.19b)
$$

\noindent
The $\eta'''$ will coincide  with (3.7), if the relation  $ - \nu_{0} d^{2} = \sigma$ is taken into account.

With the use of the expression (3.7) for  $\eta$, the current (3.5) can be brought to a tensor form:
$$
\Psi^{+} \eta \Gamma^{a} \Psi =
const \; \Psi^{+}
\left ( \begin{array}{cccc}
              +  \; \sigma  &      0      &         0      &  0  \\
                  0         &  +  g^{..}  &  -   g^{..}    &  0  \\
                  0         &  -  g^{..}  &         0      &  0  \\
                  0         &      0      &         0      &  -  H^{..,..}
\end{array}  \right ) \Gamma^{a} \Psi \; ,
$$

\noindent and further
$$
J^{a} = const \; \{ \; + \sigma \; ( C^{*} C^{a} - C C^{*a} ) -
(\Phi^{*al} \Phi_{l} - \Phi^{al} \Phi^{*}_{l} ) \} \; .
\eqno(3.20)
$$

\noindent
Since, in free case, the scalar  component  $C (x)$ vanishes (see (1.10)), the current (3.20)
will coincide with the ordinary expression for a vector particle current.
However, in presence of external electromagnetic fields, the current of generalized theory
contains a non-zero additional term proportional to the $\sigma$-factor (see (3.20)).

Lagrangian  of the theory  can be written down straightforwardly in a matrix form
$$
L = + {1 \over 2}\;\bar{\Psi}(\Gamma^{a}
\stackrel{\rightarrow}{\partial}_{a} - m) \Psi  -
{1 \over 2}\;\bar{\Psi}(\Gamma^{a} \stackrel{\leftarrow}{\partial}_{a} + m) \Psi
\eqno(3.21)
$$

\noindent with Euler-Lagrange equations as follows
$$
{\partial L \over  \partial \bar{\Psi} } -
\partial_{a} { \partial \over \partial   \bar{\Psi}_{,a} } = 0 \; , \qquad
{\partial L \over \partial  \Psi } - \partial_{a} { \partial \over \partial   \Psi_{,a} } = 0 \;
$$

\noindent having explicit forms  (3.4) and (3.1) respectively.  Tensor representation of the Lagrangian
(3.21) in basis (2.10) will follows immediately by taking into account the above block structure of all
the quantities involved. At this it is useful to have fixed some intermediate steps.
Having employed the notation
$$
[ \;C (x) , C_{l}(x),  \Phi_{l}(x), \Phi_{kl}(x) \; ] = (\Phi, C, A, F )  \; .
\eqno(3.22)
$$

\noindent we have
$$
\bar{\Psi} \Psi =
[ \;\sigma \; \Phi  ^{*} \Phi  + C^{*} g C - (A^{*} g C + C^{*} g A) - F^{*}H F \; ] \; ,
$$
$$
\bar{\Psi} \Gamma^{a} \stackrel{\rightarrow}{\partial}_{a}  \Psi =
$$
$$
= [\; \sigma \; ( - C^{*} g \Delta^{a}  \partial_{a} \Phi + \Phi ^{*} G^{a} \partial_{a} C ) -
F ^{*} H \Lambda^{a} \partial_{a} A - A^{*} g K^{a} \partial_{a} F \; ] \; ,
$$
$$
\bar{\Psi} \Gamma^{a} \stackrel{\leftarrow}{\partial}_{a}  \Psi =
$$
$$
=
[\; \sigma \; ( - C^{*} g \Delta^{a}  \stackrel{\leftarrow} {\partial}_{a} \Phi +
\Phi ^{*} G^{a} \stackrel{\leftarrow}{\partial}_{a} C ) -
F{*} H \Lambda^{a} \stackrel{\leftarrow}{\partial}_{a} A -
A^{*} g K^{a} \stackrel{\leftarrow}{\partial}_{a} F \; ]   \; ,
\eqno(3.23a)
$$

\noindent and further
$$
\bar{\Psi} \Psi  =
[ \; \sigma \; C ^{*} C +  C^{*}_{l}C^{l} -
(\Phi^{*}_{l} C^{l} + \Phi_{l} C^{*l} ) - {1 \over 2} \; \Phi^{*}_{kl} \Phi^{kl} \; ] \; ,
$$
$$
\bar{\Psi} \Gamma^{a} \stackrel{\rightarrow}{\partial}_{a}  \Psi =
[ \; + \sigma ( - C^{*l} \partial_{l} C  + C ^{*} \partial_{l} C^{l} ) -
\Phi^{*lk} \partial_{l} \Phi_{k} - \Phi^{*}_{l} \partial_{k} \Phi^{lk} \; ] \; ,
$$
$$
\bar{\Psi} \Gamma^{a} \stackrel{\leftarrow}{\partial}_{a}  \Psi =
$$
$$
=
[ \; + \sigma ( - C^{*l} \stackrel{\leftarrow}{\partial}_{l} C  +
C^{*} \stackrel{\leftarrow}{\partial}_{l} C^{l} ) -
\Phi^{*lk} \stackrel{\leftarrow}{\partial}_{l} \Phi_{k} -
\Phi^{*}_{l} \stackrel{\leftarrow}{\partial}_{k} \Phi^{lk} \; ] \; ,
\eqno(3.23b)
$$

\noindent Therefore, the Lagrangian (3.21) reads explicitly  as
$$
L = {1 \over 2} \;[  \;
- \sigma \; ( \; C^{*l} \; \partial_{l} C  + C^{l} \; \partial_{l} C^{*} ) +
 \sigma \; (C^{*} \; \partial_{l} C^{l} + C \; \partial_{l} C^{*l} \; ) \;-
$$
$$
- (\; \Phi^{*lk} \; \partial_{l} \Phi_{k} + \Phi^{lk} \; \partial_{l} \Phi^{*}_{k}\; )
- ( \;\Phi^{*}_{l} \partial_{k} \Phi^{lk} + \Phi_{l} \partial_{k} \Phi^{*lk}  \;)\; ] \; -
$$
$$
- m \; [ \; \sigma \; C^{*} C +  C^{*}_{l} C^{l} \; - \;
(\Phi^{*}_{l} C^{l} + \Phi_{l} C^{*l} ) - {1 \over 2} \;\Phi^{*}_{kl} \Phi^{kl} \; ] \; .
\eqno(3.24)
$$

\noindent
One important feature of this representation (3.24) for  $L$ is worth to be should noted specially.
The matter is that since in free case two relationships
$$
C (x) = 0 \; ,  \;\; C_{l}(x) =  + \Phi_{l}(x) \;
\eqno(3.25)
$$

\noindent hold, then the Lagrangian (3.24) reduces to the conventional Lagrangian for ordinary vector particle
$$
L^{0}  =  {1 \over 2}  \; [  \;
 -(\; \Phi^{*lk} \; \partial_{l} \Phi_{k} + \Phi^{lk} \; \partial_{l} \Phi^{*}_{k}\; ) -
( \;\Phi^{*}_{l} \partial_{k} \Phi^{lk} + \Phi_{l} \partial_{k} \Phi^{*lk}  \;)\; ] -
$$
$$
-\;  m \; [ \;  \Phi_{l} \Phi^{*l}  - {1 \over 2} \;\Phi^{*}_{kl} \Phi^{kl} \; ] \; ,
\eqno(3.26)
$$

\noindent
To prevent a possible misunderstanding  else one thing must be pointed out: going from $L$ (3.24)
to $L^{0}$ (3.26) is realized only in  the manner just described but by no means through
simple imposing $\sigma = 0$ condition.

And finally, to else one fact should be given a special  attention.  Because of a non-diagonal structure
of the invariant-form matrix $\eta$ in the basis used, the $\bar{\Psi}$-function conjugated has the following
explicit form
$$
\bar{\Psi} (x) = [\; \sigma \; C ^{*},  (C^{*l} - \Phi^{*l}),  - C^{*l}, - \Phi^{*kl} \; ] \; ,
\eqno(3.27)
$$

\noindent second and third equations in (2.11) can be found on variation of Lagrangian (3.24)
with respect to $(C^{*l} - \Phi^{*l})$ and  $(- C^{*l})$ respectively.

Going from the Lagrangian (3.21) to a  massless case is achieved by a single formal change
$$
m  \;(\; \bar{\Psi} \;\Psi \; )  \; \Longrightarrow  \;  \bar{\Psi} P \Psi \; , \;\;
P = \left ( \begin{array}{cccc}
1  &  0  &  0  &  0  \\
0  &  I  &  0  &  0  \\
0  &  0  &  0  &  0  \\
0  &  0  &  0  &  I    \end{array} \right ) \; ,
\eqno(3.28a)
$$

\noindent    from which it follows (see (3.26))
$$
- m \; [ \; \sigma \; C^{*} C +  C^{*}_{l} C^{l} -
(\Phi^{*}_{l} C^{l} + \Phi_{l} C^{*l} ) -
{1 \over 2} \;\Phi^{*}_{kl} \Phi^{kl} \; ] \Longrightarrow
$$
$$
- \; [ \; \sigma \; C^{*} C +  C^{*}_{l} C^{l}
- {1 \over 2} \;\Phi^{*}_{kl} \Phi^{kl} \; ] \; .
\eqno(3.28b)
$$

\subsection*{4.  On $C$-conjugation }

As it was shown above, the $C$-operation for the generalized field is described as a pure complex
conjugation (without any accompanying linear transformation over $\Psi$-function) in the basis (2.6).
Now we are going to see how will look $C$-operation in the initial basis (1.1).
It will be useful to consider this question in two different ways:
first,  solving a defining relation for this operation in the basis (1.1), and alternatively on
direct translating the known $C$-form in basis (2.6) into initial one.

Let  $C$-matrix in a  $\Psi (x)$-basis be known; in other words, the following relationships
$$
(\Gamma^{a} \partial_{a} - M ) \Psi (x) = 0 \; , \qquad
(\Gamma^{a} \partial_{a} - M ) \Psi^{c} (x) = 0 \; , \;
$$
$$
\Psi^{c}(x) = C \; \Psi^{*}(x) \; , \qquad   C (\Gamma^{a})^{*}C^{-1} = + \Gamma^{a} \;
\eqno(4.1)
$$

\noindent hold. Now, let $\Psi'(x)$ be associated with any new basis
$$
\Psi '(x) = S \; \Psi (x) \; ,  \qquad \Gamma^{'a} = S \Gamma^{a} S^{-1} \; ,
\eqno(4.2a)
$$

\noindent then
$$
C'\; (\Gamma^{'a})^{*}  \; (C')^{-1} = + \Gamma^{'a} \;
\eqno(4.2b)
$$

\noindent and further
$$
 S^{-1} C' S^{*} \; (\Gamma^{a})^{*}\;
S^{-1} (C')^{-1} S =  + \Gamma^{a} \qquad  \Longrightarrow  \qquad   S^{-1} C' S^{*} = C  \; .
$$

\noindent Thus, a rule for transforming  $C$-matrix is
$$
C' = S\; C \; ( S^{*})^{-1}   \; .
\eqno(4.2c)
$$

Now, let us follow up in some detail how, having  started  from  $\Psi'''$-basis (when
$C''' = I$, see (2.6)),  one can reconstruct $C$-operation in $\Psi''$-representation:
$$
\Psi '' = \left ( \begin{array}{cccc}
                         d_{0}  &  0  &  0  &  0  \\
                         0  &  d_{0}  &  0  &  0  \\
                         0  &  0  &  I  &  0  \\
                         0  &  0  &  0  &  I
\end{array} \right )
\Psi ''' \qquad   \Longrightarrow  \qquad    C'' = I  \; .
\eqno(4.3)
$$

\noindent Next step is
$$
\Psi ' = \left ( \begin{array}{cccc}
                              e^{+it}   &  0        &    0    &    0  \\
                              0         &  e^{+it}  &    0    &    0  \\
                              0         &  0        &    I    &    0  \\
                              0         &  0        &    0    &    I
\end{array} \right ) \Psi '' \;
\qquad \Longrightarrow \qquad
C' =   \left ( \begin{array}{cccc}
                        (d /d^{*})   &      0      &  0  &  0  \\
                          0          &  (d /d^{*}) &  0  &  0  \\
                          0          &      0      &  I  &  0  \\
                          0          &      0      &  0  &  I
\end{array} \right ) \; .
\eqno(4.4)
$$

\noindent  And last step is
$$
\Psi =
\left ( \begin{array}{rrrr}
               1         &     \lambda_{4}^{*}/D^{*}    &       \lambda_{2} / D^{*}     &     0   \\
               0         &     \lambda_{3}^{*} / D^{*}  &      - \lambda_{1} /D^{*}     &     0   \\
               0         &              0               &                0              &     I
\end{array} \right )
\Psi ' \qquad  \Longrightarrow \;
$$
$$
C = \left ( \begin{array}{cccc}
 d/d^{*}     &         0           &        0        &        0  \\
 0        &  [\lambda_{1}^{*} \lambda_{4}^{*} \; d  +
\lambda_{2} \lambda_{3} \; d^{*} ] / d^{*} D^{*} &
 [\lambda_{2}^{*} \lambda_{4}^{*} \;  d  -
\lambda_{2} \lambda_{4}\; d ^{*} ] / d^{*} D^{*} &  0 \\
0  & [\lambda_{1}^{*} \lambda_{3}^{*} \; d  -
\lambda_{1} \lambda_{3} \; d^{*}] / d^{*}  D^{*} &
[\lambda_{2}^{*} \lambda_{3}^{*} \; d +
\lambda_{1} \lambda_{4} \; d^{*} ] / d^{*} D^{*} & 0 \\
0  &  0  &  0  &  I
\end{array} \right ) \; .
\eqno(4.5)
$$

\noindent It should be noted that the central  two-by-two block is quite symmetric one:
$$
(\lambda_{2}^{*} \lambda_{4}^{*} \;  d  - \lambda_{2} \lambda_{4}\; d ^{*} ) =
( \lambda_{1} \lambda_{2}^{*} \lambda_{3} \lambda_{4}^{*} -
\lambda_{1}^{*} \lambda_{2} \lambda_{3}^{*} \lambda_{4} )\; ,
$$
$$
(\lambda_{1}^{*} \lambda_{3}^{*} \; d  - \lambda_{1} \lambda_{3} \; d^{*})  =
- (
\lambda_{1} \lambda_{2}^{*} \lambda_{3} \lambda_{4}^{*} -
\lambda_{1}^{*} \lambda_{2} \lambda_{3}^{*} \lambda_{4} ) \; .
\eqno(4.6)
$$

Now let us apply else one method for establishing the same $C$-matrix: via a direct  analysis of the
defining relation
$$
C (\Gamma^{a})^{*} C^{-1} = + \Gamma^{a} \qquad  \Longrightarrow \qquad
C \; (\Gamma^{a})^{*}  = \Gamma^{a} \; C \; .
\eqno(4.7a)
$$

\noindent
Taking  $C$ in the form (see (4.3))
$$
C =
\left ( \begin{array}{cccc}
                 P_{1}  &  0      &  0      &    0  \\
                 0      &  P_{2}  &  P_{3}  &    0  \\
                 0      &  P_{4}  &  P_{5}  &    0  \\
                 0      &    0    &  0      &  P_{6}
\end{array} \right ) \; ,
$$

\noindent where  $P_{1}$ is a number,   $P_{2}, P_{3},P_{4},P_{5} $  and $P_{6}$ are proportional to
a four-by-four and a six-by-six identity matrices respectively, we will have
$$
\left. \begin{array}{r}
\lambda_{1}^{*} \; P_{1} = \lambda_{1} \; P_{2}^{*} + \lambda_{2} \; P_{4}^{*} \\
\lambda_{3} \; P_{6}   = \lambda_{3}^{*} \;  P_{2}^{*} - \lambda_{4}^{*} \; P_{4}^{*}
\end{array} \right.                               \; ,
\left. \begin{array}{r}
\lambda_{2}^{*} \; P_{1}  = \lambda_{1} \; P_{3}^{*} + \lambda_{2} \; P_{5}^{*}  \\
-\lambda_{4} \; P_{6}  = \lambda_{3}^{*} \; P_{3}^{*} - \lambda_{4}^{*} \; P_{5}^{*}
\end{array} \right.                                ,
\eqno(4.8a)
$$
$$
\left. \begin{array}{r}
-\lambda_{1}^{*} \; P_{1}^{*}  = -\lambda_{1} \; P_{2} + \lambda_{2} \; P_{3}  \\
\lambda_{3} \; P_{6}^{*}  = \lambda_{3}^{*} \; P_{2} + \lambda_{4}^{*} \; P_{3}
\end{array} \right.                          \; ,
\left. \begin{array}{r}
\lambda_{2}^{*} \; P_{1}^{*}  = -\lambda_{1} \; P_{4} + \lambda_{2} \; P_{5}     \\
\lambda_{4} \; P_{6}^{*}  = \lambda_{3}^{*} \; P_{4} + \lambda_{4}^{*} \; P_{5} \; .
\end{array} \right. .
\eqno(4.8b)
$$

\noindent
One can simplify the task by taking yet known forms for  $P_{1}$ and $P_{6}$: $ P_{1} = d /d^{*}, P_{6} = 1$.
Then eqs. (4.8b) lead to
$$
\left. \begin{array}{r}
\lambda_{1}^{*} \; {d \;  \over d^{*}}  = \lambda_{1} \; P_{2}^{*} + \lambda_{2} \; P_{4}^{*} \\
\lambda_{3}    = \lambda_{3}^{*} \; P_{2}^{*} - \lambda_{4}^{*} \; P_{4}^{*}
\end{array} \right.                               \; ,
\left. \begin{array}{r}
\lambda_{2}^{*} \; { d \; \over d^{*}}   = \lambda_{1} \; P_{3}^{*} + \lambda_{2} \; P_{5}^{*}  \\
-\lambda_{4}   = \lambda_{3}^{*} \; P_{3}^{*} - \lambda_{4}^{*} \; P_{5}^{*}
\end{array} \right.                                ,
\eqno(4.9)
$$
$$
\left. \begin{array}{r}
-\lambda_{1}^{*} {d^{*} \over d\;}   = -\lambda_{1}\; P_{2} + \lambda_{2} \; P_{3}  \\
\lambda_{3}   = \lambda_{3}^{*} \; P_{2} + \lambda_{4}^{*} \; P_{3}
\end{array} \right.                          \; ,
\left. \begin{array}{r}
\lambda_{2}^{*} {d^{*} \over d\; }   = -\lambda_{1} \; P_{4} + \lambda_{2} \; P_{5}     \\
\lambda_{4}   = \lambda_{3}^{*}\;  P_{4} + \lambda_{4}^{*} \; P_{5} \;
\end{array} \right.
\eqno(4.10)
$$

\noindent and their solution follows quite straightforwardly
$$
P_{2} = { d^{*} \; \lambda_{2} \lambda_{3} +  d \; \lambda_{1}^{*} \lambda_{4}^{*} \over
(\lambda_{1} \lambda_{4}^{*} + \lambda_{2} \lambda_{3}^{*} ) d^{*}} \; , \qquad
P_{3} = { \lambda_{1}  \lambda_{2}^{*} \lambda_{3} \lambda_{4}^{*} -
\lambda_{1}^{*}  \lambda_{2} \lambda_{3}^{*} \lambda_{4}  \over
(\lambda_{1} \lambda_{4}^{*} + \lambda_{2} \lambda_{3}^{*}) d^{*} } \; ,
\eqno(4.11a)
$$
$$
P_{5} = { d \; \lambda_{2}^{*} \lambda_{3}^{*} +  d^{*} \; \lambda_{q}^{*} \lambda_{4}  \over
(\lambda_{1} \lambda_{4}^{*} + \lambda_{2} \lambda_{3}^{*} ) d^{*}} \; ,\qquad
P_{4} = { \lambda_{1}^{*}  \lambda_{2} \lambda_{3}^{*} \lambda_{4} -
\lambda_{1}  \lambda_{2}^{*} \lambda_{3} \lambda_{4}^{*}  \over
(\lambda_{1} \lambda_{4}^{*} + \lambda_{2} \lambda_{3}^{*}) d^{*} } \; ,
\eqno(4.11b)
$$

\noindent what coincides  with (4.5).

As known, a knowledge of the $C$-matrix provides us with a possibility to reconstruct
an anti-particle's   wave functions in terms of those  of particles. Also, it enables us to isolate
chargeless states $\Psi^{\pm}$ from charged ones $\Psi$:
$$
\Psi^{c} = C ^{*} \Psi = C \Psi^{*} \; , \;\;
\Psi^{\pm} = {1 \over 2} ( \Psi \pm \Psi^{c} ) \;  , \;\;
C ^{*} \Psi^{\pm}  = (\pm 1)   \Psi^{\pm}\; .
\eqno(4.12a)
$$

\noindent  Let us remember that the $U(1)$-gauge invariant derivative $(\partial_{a} - ie A_{a})$
necessarily mixes up together $\Psi^{+}$ and $\Psi^{-}$ constituents. Really,  combining two equations
$$
[\; \Gamma^{a} (\partial_{a} - ie A_{a}) - m \;] \; \Psi = 0 \; , \;
[\; \Gamma^{a} (\partial_{a} + ie A_{a}) - m \; ] \; \Psi^{c} = 0 \; , \;
$$

\noindent one can readily produce
$$
\Gamma^{a} ( \partial_{a} \Psi ^{\pm} -ig  A_{a} \Psi ^{\mp } )
 - m  \Psi^{\pm} = 0 \; .
\eqno(4.12b)
$$

\subsection*{5. Equations in a Riemannian space-time, tensor-based approach}

Section 5 deals with equations for the spin 1 particle adjusted to the general
relativity requirements. We are going  to star with  general covariant tensor formalism; later, in Sec. 7,
an alternative approach, based on the tetrad\footnote{Below, both terms, tetrad and vierbein,
will be used
interchangeably.} method by Tetrode-Weyl-Fock-Ivanenko [46-80], will
be developed in full detail. Such a study  could disclose how the additional $\sigma$-characteristic
can exhibit itself at the background of non-Eucklidean geometry. Besides,  the generally covariant
formalism will enables us to check correctness of the above determined massless limit of the theory:
reminding that  massless particle's covariant equations are assumed to be  conformally invariant.

On taking a simple formal change of   Cartesian  derivatives and tensors  into  covariant ones
$$
\partial_{b}  \Longrightarrow  \nabla_{\beta} \;, \;
C_{b}(x)  \Longrightarrow C_{\beta}(x)  \; , \;
\Phi_{b} (x) \Longrightarrow    \Phi_{\beta} (x) \; , \;
\Phi_{a b} (x) \Longrightarrow    \Phi_{\alpha \beta}(x) \;
$$

\noindent eqs. (2.6) take on the form
$$
\nabla^{\alpha} C_{\alpha} -  m   C = 0  \; , \;
\nabla^{\alpha} \Phi _{ \beta \alpha }  -  m  C_{\beta} = 0  \; ,
$$
$$
\sigma \; \nabla_{\beta} C +
\nabla^{\alpha} \Phi_{\beta \alpha } -   m \Phi_{\beta} = 0 \; , \;
$$
$$
\nabla_{\alpha} \Phi_{\beta} - \nabla_{\beta} \Phi_{\alpha}  -
 m  \Phi _{\alpha \beta} = 0 \; .
\eqno(5.1)
$$

\noindent
Obviously, in presence of external electromagnetic fields, with the notation
$ \nabla _{\alpha} \Longrightarrow D_{\alpha} =  \nabla_{\alpha} -i g   A_{a}(x)  \; $,
instead of (5.1) we will have
$$
D^{\alpha} C_{\alpha} -  m C = 0  \; , \;
D^{\alpha } \Phi _{ \beta \alpha } - m   C_{\beta} = 0 \; ,
$$
$$
\sigma \; D_{\beta} C +
D^{\alpha} \Phi_{\beta \alpha  } - m  \Phi_{\beta} = 0 \; , \;
$$
$$
D_{\alpha} \Phi_{\beta} - D_{\beta} \Phi_{\alpha} -  m
\Phi _{\alpha \beta} = 0\;.
\eqno(5.2)
$$

 Let us exclude the field $C$ from third equation in (5.2). To this end,
one can produce
$$
D^{\alpha}  C_{\alpha } = {1  \over m}
D^{\alpha }  D^{\beta } \Phi _{ \alpha \beta }  \; , \qquad
D ^{\alpha } D^{\beta} \Phi_{\alpha \beta   }
=  \nabla ^{\alpha } \nabla^{\beta} \Phi_{\alpha  \beta } -
 i g \; {1 \over 2} \; F^{\alpha \beta} \Phi_{\alpha  \beta} \; ,
$$

\noindent  where $F_{\alpha \beta } = ( \nabla_{\alpha} A_{\beta} - \nabla_{\beta} A_{\alpha})$,
and further
$$
\nabla ^{\alpha } \nabla^{\beta} \Phi_{\alpha \beta } =
{1 \over 2} \;  ( \nabla ^{\alpha } \nabla^{\beta} - \nabla ^{\beta }
 \nabla^{\alpha} )   \Phi_{\alpha \beta  } =
$$
$$
= {1\over 2} [\;
\Phi _{\nu \alpha }(x) R^{\nu \;\; \beta \alpha }_{\;\;\beta }(x) +
\Phi _{\beta \nu }(x) R^{\nu \;\;  \beta \alpha }_{\;\;\alpha }(x)\; ] =
{1\over 2} [\;   \Phi _{ \nu \alpha }(x) R^{\nu \alpha }(x) +
\Phi _{\nu \beta }(x) R^{\nu \beta }(x)\; ] \; .
$$

\noindent Now, remembering symmetry of the Ricci tensor  $R_{\alpha \beta }(x)$,
we arrive at
$$
D^{\alpha }  C_{\alpha}  = +i \;  { g  \over   m } \;
{1 \over 2 } \; F^{ \alpha \beta  } \Phi_{\alpha \beta} \; .
$$

\noindent Since,  first equation in (5.2) yields
$$
C = - i    \;  { g \over  m^{2} } \; {1 \over 2}  F^{\alpha \beta } \Phi_{\alpha \beta} \; .
\eqno(5.3)
$$

\noindent Therefore, the main equations in (5.2) read as
$$
D^{\alpha}   C_{\alpha} -  m  C = 0  \; , \qquad
D^{\alpha }  \Phi _{ \beta \alpha } - m  C_{\beta} = 0 ,
$$
$$
- i \sigma \;  { g   \over   m^{2}  } \;
D_{\rho} \; ( { 1 \over 2} F^{\alpha \beta} \Phi_{\alpha \beta} ) +
D^{\alpha}  \Phi_{\rho \alpha  }  - m  \Phi_{\rho} = 0 \; ,
$$
$$
D_{\alpha} \Phi_{\beta} - D_{\beta}  \Phi_{\alpha} -  m
 \Phi _{\alpha \beta} = 0 \; .
\eqno(5.4)
$$

\noindent
In absence external electromagnetic fields, the term proportional to $\sigma$
vanishes. This means that $\sigma$-characteristics manifests itself physically
only in electromagnetic fields but not in gravitational ones.

Now, let us consider a generally covariant massless equations. They are to be as follows
$$
\nabla^{\alpha} C_{\alpha} -  C = 0  \; , \qquad
\nabla^{\alpha } \Phi _{ \beta \alpha } -   C_{\beta} = 0  \; ,
$$
$$
\sigma \; \nabla_{\beta} C +
\nabla^{\alpha} \Phi_{\beta \alpha }  = 0 \; , \qquad
\nabla_{\alpha} \Phi_{\beta} - \nabla_{\beta} \Phi_{\alpha} -
\Phi _{\alpha \beta} = 0\;.
\eqno(5.5)
$$

As in the flat space-time, here we have
$$
C = 0  \; , \qquad C_{\beta} = 0 ,
$$
$$
\nabla^{\alpha} \Phi_{\beta \alpha  } = 0 \; , \qquad
\nabla_{\alpha} \Phi_{\beta} - \nabla_{\beta} \Phi_{\alpha} -
\Phi _{\alpha \beta} = 0\; .
\eqno(5.6)
$$

Thus, eqs. (5.6) happen to be absolutely equivalent to ordinary generally covariant Proca  massless
equations.

Repeating all arguments  after formula (1.18)  from Sec. 1,  for a generally covariant massless particle
in an external vector field, we   will get
$$
C_{\beta} =  D^{\alpha}  \Phi_{\alpha \beta } \; , \qquad
C =   -i g  \; {1 \over 2} \;  \; F^{\alpha \beta } \Phi_{\alpha \beta } \; ,
$$
$$
- i \; \sigma  g  \;  D_{\rho}
( {1 \over 2 }\; F^{\alpha \beta }\Phi_{\alpha \beta }) +
D^{\alpha} \Phi_{\rho \alpha } = 0 \; , \;
$$
$$
D_{\alpha}  \Phi_{\beta} - D_{\beta} \Phi_{\alpha} -
\Phi _{\alpha \beta} = 0\; .
\eqno(5.7)
$$

\subsection*{6. Covariant tetrad-based  formalism }

We start with the matrix equation in the Minkowski space-time  (2.12)
$$
(\;   \Gamma^{a} \; \partial_{a} \; - \; m \; )\;  \Psi  (x) = 0 \; ,
\eqno(6.1a)
$$
$$
\Psi  = ( C, C_{a} , \Phi _{a} , \; \Phi _{ab} ) , \;
\Gamma^{a} =
\left ( \begin{array}{llll}
0  &  G^{a} &  0  &  0  \\
0  &  0  &  0  &  K^{a} \\
\sigma \Delta^{a}  &  0  &  0 &  K^{a} \\
0 &  0  &  \Lambda ^{a}  &  0
\end{array} \right ) \; ,
\eqno(6.1b)
$$
$$
(K^{a}) _{l} ^{\;\;mn} =
(- g^{am} \delta^{n}_{l} + g^{an} \delta^{m}_{n}) \; , \;
$$
$$
(\Lambda^{a})_{bl}^{\;\;\;\;k} = \delta_{bl}^{ak}  \; , \;
(G^{a})_{(0)}^{\;\;\;k} = g^{ab} \; , \;\;
(\Delta^{a})^{\;\;\;(0)}_{l} =  \delta^{a}_{l} \; .
\eqno(6.1c)
$$

In accordance with the known procedure by Tetrode-Weyl-Fock-Ivanenko,
eq.  (6.1) is extended to a curved space-time with a metric tensor $g_{\alpha \beta }(x)$
and an accompanying tetrad  $e^{\alpha }_{(a)}(x) $  as follows
$$
[ \; \Gamma ^{\alpha }(x)\; ( \partial_{\alpha} \;  +  \;
B_{\alpha }(x) ) \; - m \;  ] \;\Psi  (x)  = 0 \; ,
\eqno(6.2)
$$

\noindent where
$$
\Gamma ^{\alpha }(x) = \Gamma ^{a} e ^{\alpha }_{(a)}(x) \; , \qquad
B_{\alpha }(x) =  {1 \over 2}\; J^{ab} e ^{\beta }_{(a)}\nabla _{\alpha }( e_{(b)\beta }) \; .
$$

\noindent Here $J^{ab}$ designate generators of the Lorentz group representation (see below formulas (6.6))
associated with the set $(C, C_{k}, \Phi_{k},  \Phi_{kl})$.

The equation (6.2) involves the object $e^{\alpha }_{(a)}(x)$. Therefore,  similar to the case of much more
familiar spin 1/2 particle,   there must exist possibility  that any such equations associated respectively
with $e^{\alpha }_{(a)}(x)$   and    $e^{'\alpha }_{(a)}(x)$ were translated into each other; otherwise eq.
(6.2) is incorrect.

Let us show that if two tetrads are connected  by a Lorentz operator $L$:
$$
e^{\alpha }_{(a)}(x)  \; \Longrightarrow   \;
e'^{\alpha }_{(b)}(x)\; = \;  L^{\;\;b}_{a} (x) \; e^{\alpha }_{(b)}(x)\; ,
\eqno(6.3a)
$$

\noindent   the respective two equations
$$
[\;   \Gamma ^{\alpha }(x) (\partial_{\alpha}  + B_{\alpha }(x)) \;  -
\; m \; ] \; \Psi  (x)  = 0 \; , \;
$$
$$
[\;  \Gamma'^{\alpha }(x) (\partial_{\alpha}  + B'_{\alpha }(x))  -
m \; ]\;  \Psi'(x)  = 0 \;
\eqno(6.3b)
$$

\noindent can be set into each other  by means of a Lorentz gauge transformation
$$
\Psi ' (x) = S(x)\; \Psi (x)     \; ,\;\;
\left ( \begin{array}{c}
C' (x)  \\  C'_{k}(x) \\   \Phi'_{k}(x) \\ \Phi'_{kl}(x)
\end{array} \right ) =
\left ( \begin{array}{cccc}
1 &  0  &  0  &  0  \\
0 &  L_{k}^{\;\;l} & 0  &  0  \\
0 &  0  &  L_{k}^{\;\;l} & 0  \\
0 &  0  &  0  &  L_{k}^{\;\;m} L_{l}^{\;\;n}
\end{array} \right ) \;\;
\left ( \begin{array}{l}
C (x) \\ C_{l}(x)  \\    \Phi_{l}(x) \\ \Phi_{mn}(x)
\end{array} \right )                \; .
\eqno(6.3c)
$$

It is useful time to dwell upon some peculiarities of the formalism.  The form of the $15 \times 15$-gauge
transformation (6.3c) by no means takes into consideration antisymmetry of the  tensor $\Phi_{kl}$.
If such a symmetry is allowed for, then the Lorentz operator can be rewritten in a modified form
$$
\Phi'_{kl}(x) = L_{k}^{\;\;m} L_{l}^{\;\; n} \Phi _{mn} =
{1 \over 2 } ( L_{k}^{\;\;m} L_{l}^{\;\; n} - L_{l}^{\;\;m} L_{k}^{\;\; n}   ) \; \Phi _{mn} \; .
\eqno(6.4a)
$$

\noindent This formula (6.4a) has some advantage  in the context of the matrix formalism, in view of
presence of the tensor $\Phi_{mn}$ just as a 6-dimension object
$(\Phi_{01},\Phi_{02}, \Phi_{03},\Phi_{23},\Phi_{31},\Phi_{12} )$. At this
the 6-vector transforms by means of the operator
$$
S_{6\otimes 6} = S_{kl}^{\;\;\;\;mn} =
 ( L_{k}^{\;\;m} L_{l}^{\;\; n} - L_{l}^{\;\;m} L_{k}^{\;\; n}   ) \; , \;
или \;\; S_{6 \otimes 6 } = L _{[.} \otimes L_{.]} \; .
\eqno(6.4b)
$$

\noindent The property noted plays a role and for corresponding generators. Let $V^{ab}$ and
$(V \otimes V) ^{ab}$ designate respectively vector and tensor generators:
$$
\Phi _{s}'  = (\delta_{s}^{p} + {1 \over 2} \delta \omega_{ab} \; (V^{ab})_{s}^{\;\;p} )
 \Phi_{p} \; , \;\;  (V^{ab})_{s}^{\;\; p}  = - g^{ap} \delta_{s}^{b} + g^{bp} \delta^{a}_{s} \; ,
\eqno(6.5a)
$$
$$
\Phi_{mn}' = ( (\delta_{m}^{s} + {1 \over 2} \delta \omega_{ab} \; (V^{ab})_{m}^{\;\;s} )
(\delta_{n}^{p} + {1 \over 2} \delta \omega_{ab} \; (V^{ab})_{n}^{\;\;p} ) \Phi _{sp} =
$$
$$
= \delta_{m}^{s} \delta_{n}^{p} + {1 \over 2} \delta \omega_{ab} \;
[(V \otimes V)^{ab}]_{mn}^{\;\;\;\;sp} ] \Phi_{sp} \; ,
$$

\noindent where
$$
[(V \otimes V)^{ab}] _{mn}^{\;\;\;\;sp} =
[ (-g^{as}  \delta^{b}_{m}  +       g^{bs}  \delta_{m}^{a})   \delta_{n}^{p} +
\delta_{m}^{s} (- g^{ap} \delta^{b}_{n}  +  g^{bp} \delta_{n}^{a}   ) \; ] \; ,
\eqno(6.5b)
$$

\noindent or
$$
(V \otimes V)^{ab} = V^{ab} \otimes I + I \otimes V^{ab} \; .
$$

\noindent
Here the tensor  generator does not take into consideration any symmetry property of  $\Phi_{sp}$,
otherwise we will have
$$
[V^{ab}_{[\;]}] _{mn}^{\;\;\;\;sp} \;\; \Phi_{sp} =
{1 \over 2} \{ \; [(V \otimes V)^{ab}] _{mn}^{\;\;\;\;sp} -
                  [(V \otimes V)^{ab}] _{mn}^{\;\;\;\;ps} \} \Phi_{sp} =
$$
$$
= {1 \over 2} \; \{ ( - \delta^{as}_{mn} g^{bp} +  \delta_{mn}^{ap} g^{bs} ) -
     ( - \delta^{bs}_{mn} g^{ap}  + \delta_{mn}^{bp} g^{as} ) \}
\Phi_{sp} \; .
\eqno(6.5в)
$$

Thus, there exist two usages of the generators $J^{abn}$ and gauge transformation $S$:
$$
J^{ab} = \left ( \begin{array}{cccc}
0  &  0  &  0  &  0  \\
0  & V^{ab}  &  0  &  0  \\
0  &  0   &  V^{ab}  & 0  \\
0  &  0 &  0 &  (V \otimes V)^{ab}
\end{array} \right ) \; , \;
S = \left ( \begin{array}{cccc}
1  &  0  &  0  &  0 \\
0  &  L  &  0  &  0  \\
0  &  0  &  L  &  0 \\
0  &  0  &  0  & L \otimes L
\end{array} \right )     \; ,
\eqno(6.6a)
$$

\noindent or
$$
J^{ab}_{[]} = \left ( \begin{array}{cccc}
0  &  0  &  0 & 0  \\
0  & V^{ab}  &  0  &  0  \\
0  &  0  &  V^{ab}  & 0  \\
0  &  0 &  0 &  (V_{[\;]} )^{ab}
\end{array} \right ) \; , \;
S_{[\;]} = \left ( \begin{array}{cccc}
1  &  0  &  0  &  0 \\
0  &  L  &  0  &  0  \\
0  &  0  &  L  &  0 \\
0  &  0  &  0  & L_{[.} \otimes L_{.]}
\end{array} \right )   \;
\eqno(6.6b)
$$

Now, starting from an equation for $\Psi$-function, let us obtain an equation for
$\Psi'$-function.  We have
$$
\left [ \;  S  \Gamma ^{\alpha }  S^{-1}
 \left  (\partial _{\alpha}  +
S \; B_{\alpha }\; S^{-1} +  S  \partial_{\alpha} S^{-1}
\right ) \; -
 \; M \;\right ] \; \Psi' = 0  \; ,
$$

\noindent and two relationships are to be proved
$$
S \; \Gamma ^{\alpha } \; S^{-1} =
\Gamma'^{\alpha}      \; ,
\eqno(6.7a)
$$
$$
 S \; B_{\alpha} \; S^{-1} \; + \; S \; \partial_{\alpha}\; S^{-1}
 = B'_{\alpha}     \; .
\eqno(6.7b)
$$

\noindent The first  (6.7a)  can be rewritten as
$$
S \; \Gamma ^{a} \; e ^{\alpha }_{(a)} \; S^{-1} \; =
\Gamma ^{b} \; e'^{\alpha }_{(b)}   \; \Longrightarrow \;
S \;\Gamma ^{a} \; S^{-1} \; = \; \Gamma^{b} \; L^{\;\;a}_{b} \; .
\eqno(6.8a)
$$

\noindent The last is a well-known condition insuring the relativistic invariance property  of the
15-component wave equation being considered in the Minkowski space. In blocks it reads
$$
G^{a} \; L^{-1}  = G^{b} L_{b}^{\;\;a} \; , \;\;
L \; \Delta^{a} = \Delta^{b} \; L_{b}^{\;\;a} \; ,
$$
$$
L \; K^{a} ( L^{-1} \otimes L^{-1}) = K^{b} \; L_{b}^{\;\;a} \; , \;
(L \otimes L) \Lambda^{a} \; L^{-1} = \Lambda^{b} \; L_{b}^{\;\;a} \; .
\eqno(6.8b)
$$

\noindent One can check these relations by a direct calculation
with the use of the explicit form of all the  blocks  $G^{a}, \Delta^{a}, K^{a}, \Delta^{a}$.
The first one gives
$$
(G^{a})_{(0)}^{\;\;\; k} (L^{-1})_{k}^{\;\;c} = (G^{b})_{(0)}^{\;\;\;c} L_{b}^{\;\;a}
\;\; \Longrightarrow \;\; (L^{-1})^{ac} = L^{ca} \; ,
$$

\noindent which coincides with the known pseudo orthogonality property of the Lorentz matrix.
Second one leads to an identity
$$
L_{k}^{\;\; l} (\Delta^{a})_{l}^{\;\;(0)}  = (\Delta^{b})_{k}^{\;\;(0)} L_{b}^{\;\;a}
\;\;\Longrightarrow \;\; L_{k}^{\;\;a} = L_{k}^{\;\;a} \; ;
$$

\noindent Third yields
$$
L_{d}^{\;\;l} (K^{a})_{l}^{\;\;mn} [  (L^{-1}) _{m}^{\;\;s} (L^{-1})_{n}^{\;\;p}) =
(K^{b}) _{d}^{\;\;sp} L_{b}^{\;\;a} \; ,
$$

\noindent  what is equivalent to
$$
\delta^{p}_{d} (L^{-1})^{as} - \delta^{s}_{d} (L^{-1})^{ap} =
\delta^{p}_{d} \; L^{sa} - \delta^{s}_{d} \;  L^{pa} \; .
$$

\noindent And finally forth reads as
$$
L_{m}^{\;\;c}  L_{n}^{\;\;d} \; ( \Lambda^{a} ) _{cd}^{\;\;\;\;k} \; (
L^{-1}) _{k}^{\;\;l} = \Lambda^{b} \; L_{b}^{\;\;a}
$$

\noindent
that is an identity
$$
\delta_{n}^{l} L_{m}^{\;\;a} - \delta_{m}^{l} L_{n}^{\;\;a} =
\delta_{n}^{l} L_{m}^{\;\;a} - \delta_{m}^{l} L_{n}^{\;\;a} \;  .
$$

\noindent
Now let us turn to eq. (6.7b) and consider in some detail the term
$$
S \; B_{\alpha} \; S^{-1} =  {1 \over 2} \;
J^{ab}   S^{-1} \;  e_{(a)}^{\;\beta} (\nabla_{\alpha} e_{(b)\beta})   \; .
\eqno(6.9)
$$

\noindent Expressing  the vierbein $e_{(b)\beta}$  in terms of the primed $e'_{(b)\beta} $, we get
$$
S \; B_{\alpha} \; S^{-1} = {1 \over 2} \;( S \; J^{ab} S^{-1} )
\; (L^{-1})_{a}^{\;\;k} e_{(k)}^{'\;\beta} \;
[ \; \nabla_{\alpha} (L^{-1})_{b}^{\;\;l} e_{(l)\beta} \; ]
$$

\noindent and further
$$
S \; B_{\alpha} \; S^{-1} = {1 \over 2} \;( S \; J^{ab} S^{-1} ) \times
$$
$$
\times \; \; \left [ \;  (L^{-1})_{a}^{\;\;k} e_{(k)}^{'\;\beta }\;
\; [ \; ( \partial _{\alpha} (L^{-1})_{b}^{\;\;l} \; )  \;
e'_{(l)\beta}  +  (L^{-1})_{b}^{\;\;l} \nabla_{\alpha} e'_{(l)\beta} \; \right ] =
$$
$$
=
{1 \over 2} \;( S \; J^{ab} S^{-1} ) \;  \left [
 \; (L^{-1})_{a}^{\;\;k} \;
 (\partial _{\alpha} (L^{-1})_{b}^{\;\;l})\;
e_{(k)}^{'\;\beta} \;  e'_{(l)\beta} \;  ] +
(L^{-1})_{a}^{\;\;k}  (L^{-1})_{b}^{\;\;l} \;
e_{(k)}^{'\;\beta} \;  \nabla_{\alpha} e'_{(l)\beta} \right ] \; .
$$

\noindent So, one arrives at
$$
S \; B_{\alpha} \; S^{-1} = {1 \over 2} \;( S \; J^{ab} S^{-1} ) \; \times
$$
$$
\times \;
\left [
(L^{-1})_{a}^{\;\;k}
(\partial_{\alpha} (L^{-1})_{b}^{\;\;l})\; g_{kl} +
(L^{-1})_{a}^{\;\;k}  (L^{-1})_{b}^{\;\;l} \;
e_{(k)}^{'\;\beta} \;  \nabla_{\alpha} e'_{(l)\beta} \right ]  \; .
\eqno(6.10)
$$

\noindent Taking into account the identity (its proof  will be given below)
$$
S J^{ab} S^{-1} = J^{mn} L_{m}^{\;\;\;a} L_{n}^{\;\;b} \; ,
\eqno(6.11)
$$

\noindent  from (6.10) one readily obtains
$$
S \; B_{\alpha} \; S^{-1} =
{1 \over 2} \; J^{mn} L_{n}^{\;\;b} \partial_{\alpha}
(L^{-1})_{bm} +   { 1 \over 2} \;
J^{kl} e_{(k)}^{'\;\beta} \;  \nabla_{\alpha} e'_{(l)\beta} =
$$
$$
= {1 \over 2} \; J^{mn} L_{n}^{\;\;b}
\partial _{\alpha} (L^{-1})_{bm}  +  B'_{\alpha}  \; .
\eqno(6.12)
$$

\noindent Now, on taking into consideration (6.12),  from  (6.7b ) it follows
$$
S \partial_{\alpha} S^{-1}  +   {1 \over 2} \; J^{mn}
L_{n}^{\;\;b} \partial _{\alpha} (L^{-1})_{bm} = 0 \;
$$

\noindent  which  can be rewritten as
$$
S \partial_{\alpha} S^{-1} = {1 \over 2} \; J^{mn}
 L_{mb} \; ( \partial _{\alpha} L_{n}^{\;\;b} )   \; .
\eqno(6.13)
$$

\noindent With the use of all blocks involved, the previous relation  will read as
$$
L \partial_{\beta} L^{-1} =
{1 \over 2} V^{ab}  L_{ad} \partial _{\beta} L_{b}^{\;\;d} \; ,
\eqno(6.14а)
$$
$$
(L ^{-1}\otimes L^{-1} ) \partial _{\beta} ( L ^{-1}\otimes L^{-1})  =
{1 \over 2} (V \otimes V)^{ab}  L_{ad} \partial _{\beta} L_{b}^{\;\;d}  \; .
\eqno(6.14b)
$$

\noindent It suffices to prove only  (6.14a), because eq. (6.14b) is a straightforward result from
that on using  the formula $(V \otimes V)^{ab} = (V^{ab} \otimes I + I \otimes V^{ab} )$.
Eq.  (6.14a)  yields
$$
L _{s}^{\;\;k} \partial_{\beta} (L^{-1})_{k}^{\;\;p} =
{1 \over 2} (g^{ap} \delta_{s}^{b} - g^{bp} \delta_{s}^{a})
\;   L_{ad} \partial _{\beta}  L_{b}^{\;\;d} \; ,
$$

\noindent  or
$$
L _{sd} \partial_{\beta} (L^{-1})^{dp} = {1 \over 2} \;
[ L^{pd} \partial_{\beta} L_{sd} - L_{sd} \partial_{\beta} L^{pd} ] \; .
$$

\noindent Here a supplementary relation is needed, which  may be resulted in by differentiating the known
pseudo-orthogonality relation for $L$:
$$
\partial_{\beta}\;  [ \; L_{sd}  L^{pd} \;] = ( \partial_{\beta} L_{sd} ) \;  L^{pd}  +
L_{sd} \; \partial_{\beta} L^{pd}    = 0 \;
\eqno(6.15)
$$

\noindent  With the use of eq. (6.15) the previous one reads as identity. So, eq. (6.13) has been proved.  .

Now, we are to return to eq. (6.11) used above. It is equivalent to
$$
L V^{ab} L^{-1} = V^{mn} \; L_{m}^{\;\;\;a} L_{n}^{\;\;b} \; , \;
\eqno(6.16a)
$$
$$
(L \otimes L ) (V \otimes V) (L^{-1} \otimes L^{-1}) =
(V \otimes V)^{mn} \; L_{m}^{\;\;\;a} L_{n}^{\;\;b} \; .
\eqno(6.16b)
$$

\noindent    Obviously, the second is a simple consequence of the first. In turn, the first one
(6.16a) happens to be an identity:
$$
L (V^{ab})_{s}^{\;\;p} L^{-1} = (V^{mn})_{s}^{\;\;p}
 \; L_{m}^{\;\;\;a} L_{n}^{\;\;b} \Longrightarrow
$$
$$
- L_{k}^{\;\;b} L^{sa} + L_{k}^{\;\; a} + L_{k}^{\;\;a} L^{sb} =
- L_{k}^{\;\;b} L^{sa} + L_{k}^{\;\; a} + L_{k}^{\;\;a} L^{sb} \; .
$$

Thus, any two equations (6.3б) associated with two different tetrads
are transforms of each other. This means that the  gauge invariance principle with respect to
local Lorentz group, in a manner like more familiar case of spin 1/2 particle, holds.

Else one fact should be mentioned. The wave equation under consideration will look as correct one
in a generally  relativistic context, if the 15-component wave function $\Psi(x)$ is a scalar with
respect to general coordinate transformations:
$$
x'^{\alpha } =
f^{\alpha }(x^{\alpha}) \; \Longrightarrow \; \Psi'(x') = \Psi (x) \; .
\eqno(6.17)
$$

Now let us show --- how tensor equations discussed in Sec. 5 are resulted in from the matrix one:
$$
[\; \Gamma^{\rho}(x) ( \partial + B_{\rho}(x) )   - m \; ]\; \Psi (x) = 0 \; .
\eqno(6.18)
$$

\noindent
To this end, as a first step, let us take into consideration the block structure of all objects involved:
$$
\Gamma^{\rho}(x) = e^{\rho}_{(l)}  \Gamma^{l}=
\left ( \begin{array}{cccc}
0  &  G^{\rho}  &  0  &  0  \\
0  &  0  &  0  &  K^{\rho} \\
\sigma \Delta^{\rho}  &  0  &  0  &  K^{\rho} \\
0  &  0  &  \Lambda^{\rho}  &  0
\end{array} \right )\; , \;
B_{\rho}(x) = \left ( \begin{array}{cccc}
0  &  0  &  0 & 0  \\
0  &  V_{\rho}   & 0  &  0  \\
0  &  0  &  V_{\rho}  &  0 \\
0  &  0  &  0  &  (V \otimes V)_{\rho}
\end{array} \right ) \; , \;
$$
$$
\Gamma^{\rho} (\partial_{\rho} + B_{\rho}) =
\left ( \begin{array}{cccc}
0   &  G^{\rho} [\partial_{\rho} + V_{\rho} ]  &  0  &  0  \\
0  &  0  &  0  &  K^{\rho} [\partial_{\rho} + (V \otimes V)_{\rho} ] \\
\sigma \Delta^ {\rho}  \partial_{\rho} &  0  &  0  &
K^{\rho} [ \partial_{\rho} + (V \otimes V)_{\rho} ] \\
0  &  0  &  \Lambda^{\rho} [ \partial_{\rho} + V_{\rho} ] &  0
\end{array} \right ) \; .
\eqno(6.19)
$$

\noindent Below it will convenient to  employ the notation
$$
(V_{\rho})_{l}^{\;\;k}(x)  =
 {1 \over 2 } (V^{ab})_{l}^{\;\;k} \; e_{(a)}^{\beta}  \nabla_{\rho} e_{(b)\beta} \; , \;
\eqno(6.20)
$$
$$
[(V \otimes V)_{\rho}]_{ps}^{\;\;\;\;mn} (x) =
 {1 \over 2 }  [ (V \otimes V)^{ab} ]_{ps}^{\;\;\;\; mn} \;
e_{(a)}^{\beta} \nabla_{\rho} e_{(b)\beta} \; .
$$

\noindent
Then eq.(6.18) takes on the block form
$$
[\;  G^{\rho} \; ( \partial_{\rho} \; + \; V_{\rho} ) ]_{(0)}^{\;\;\;k} \; C_{k}  = m C  \; ,
$$
$$
\{
K^{\rho} \; [ \partial_{\rho}  \;+ \; (V \otimes V)_{\rho} ] \}_{l}^{\;\;mn} \;  \Phi_{mn}  = m C_{l} \; ,
$$
$$
\sigma \;  (\Delta^{\rho})_{(l)}^{\;(0)} \; \partial_{\rho} \;   C  \; + \;
\{\;  K^{\rho}\; [ \; \partial_{\rho} + (V \otimes V)_{\rho} \; ] \;  \}_{l}^{\;\;mn}  \;
\Phi_{mn}  = m \Phi_{l} \; ,
$$
$$
[\; \Lambda^{\rho} \;  ( \partial_{\rho} \; + \;  V_{\rho}) ]_{mn}^{\;\;\;\;k}\;
 \Phi_{k} = m \Phi_{mn} \; .
\eqno(6.21)
$$

\noindent With the use of explicit form of the all block-matrices in (6.21) one can readily arrive at
$$
e^{(k)\alpha} \partial_{\alpha} C_{k} + \gamma_{l}^{\;\;kl}  \; C_{k} = m C \; ,
$$
$$
e^{(k)\alpha} \partial_{\alpha} \Phi_{lk} + \gamma_{l}^{\;\;mn} \Phi_{mn} +
\gamma_{d}^{\;\;kd} \Phi_{lk} = m C_{l} \; ,
$$
$$
\sigma e^{\alpha}_{(l)} \partial_{\alpha} C +
e^{(k)\alpha} \partial_{\alpha} \Phi_{lk} + \gamma_{l}^{\;\;mn} \Phi_{mn} +
\gamma_{d}^{\;\;kd} \Phi_{lk} = m \Phi_{l} \; ,
$$
$$
e^{\alpha}_{(m)} \partial_{\alpha} \Phi _{n} -
e^{\alpha}_{(n)} \partial_{\alpha} \Phi _{m}  +
(\gamma^{k}_{\;\;mn} -  \gamma^{k}_{\;\;nm} ) \Phi_{k} = m \Phi_{mn} \; .
\eqno(6.22)
$$

\noindent
Here  $\gamma_{abc}$  designates the Ricci rotation coefficients
$$
\gamma_{abc} (x) = - e_{(a)\rho ; \sigma} e_{(b)}^{\rho}  e_{(c)}^{\sigma}  \; .
$$

\noindent In turn, as could be checked,  eqs. (6.22) represent  the following generally
covariant tensor equations (just those introduced in Sec. 5)
$$
\nabla^{\alpha} C_{\alpha} - m   C = 0  \; , \;\;
\nabla^{\alpha} \Phi _{ \beta \alpha  }  - m   C_{\beta} = 0 ,
$$
$$
\sigma \; \nabla_{\beta} C +
\nabla^{\alpha} \Phi_{\beta \alpha } -  m  \Phi_{\beta} = 0 \; , \;
\nabla_{\alpha} \Phi_{\beta} - \nabla_{\beta} \Phi_{\alpha}  -
m   \Phi _{\alpha \beta} = 0 \; ,
\eqno(6.23)
$$

\noindent in the  vierbein  form. Connection between generally covariant and  tetrad representatives of
the wave function is determined quite conventionally
$$
C_{\alpha}  = e_{\alpha}^{(l)} C_{l} \; ,
\Phi_{\alpha}  = e_{\alpha}^{(l)} \Phi_{l} \; , \;
\Phi_{\alpha \beta }  = e_{\alpha}^{(m)} e_{\alpha}^{(n)} \Phi_{mn} \; .
\eqno(6.24)
$$

\subsection*{7.  Bilinear invariants in a curved space-time }

This Section deals with a generally covariant equation for a $\bar{\Psi}$-function
conjugate to the $\Psi$ and methods for constructing some bilinear invariants in terms of
$\bar{\Psi}$ and $\Psi$. A  knowing of this matter will enable us to produce  automatically
expressions for a  Lagrangian and  a conserved  current (about the energy-momentum tensor see Sec. 9).

The basic  requirement of an invariance matrix $\eta$ is that a form  $\Psi^{+} \eta \Psi$ be
invariant under Lorentz group transformations
$$
\Psi^{+} \eta \Psi = inv   , \qquad \Psi' = S \Psi \;  ,
$$

\noindent which necessitates
$$
\eta = S^{+} \eta S \qquad  \Longrightarrow \qquad \eta S^{-1} = S^{+} \eta \; .
\eqno(7.1a)
$$

\noindent Being taken for an infinitesimal $S$,   eq. (7.1a) reads
$$
- \eta \; J^{ab} = (J^{ab})^{+}  \eta \; ;
\eqno(7.1b)
$$

\noindent it is what we need. Now, starting from  an equation for $\Psi$
$$
[ \; \Gamma^{\alpha}  (\partial_{\alpha} + B_{\alpha}   ) - m \; ] \; \Psi = 0 \; ,
\eqno(7.2)
$$

\noindent we are going to reconstruct that for  $\bar{\Psi} = \Psi^{+} \; \eta $.
From eq. (7.2) it follows
$$
\Psi^{+}\eta \eta^{-1} [ \; (\stackrel{\leftarrow}{\partial}_{\alpha}
+ B_{\alpha}^{\;+})  \; \eta  \eta^{-1} \; (\Gamma^{\alpha})^{+}  - m \; ] \; \eta = 0     \; ;
$$

\noindent which yields
$$
\eta^{-1} \; [\Gamma^{\alpha}(x)]^{+} \; \eta = - \Gamma^{\alpha}(x) \; , \;\;
\eta^{-1} [B_{\alpha}(x)]^{+} \eta = - B_{\alpha}(x) \; ,
\eqno(7.3)
$$

\noindent and further (compare with (7.2))
$$
\bar{\Psi}\;  [\; ( \stackrel{\leftarrow}{\partial}_{\alpha}
- B_{\alpha} ) \;  \Gamma^{\alpha}   + M \; ] \; = 0     \; .
\eqno(7.4)
$$

\noindent Take notice on the sign 'minus' at the spinor connection $B_{\alpha}$ in (7.4).
Now, multiplying eq. (7.2) from the left by $\bar{\Psi}$, and eq. (7.4) --- from the right by
$\Psi $, and adding results together, we get to
$$
\bar{\Psi} \stackrel{\leftarrow}{\partial}_{\alpha}   \; \Gamma^{\alpha} \Psi  \; +  \;
\bar{\Psi}  \Gamma^{\alpha} \; \stackrel{\rightarrow}{\partial}_{\alpha}
\Psi \; +
$$
$$
+ \;\bar{\Psi}  \; ( \Gamma^{\alpha}  B_{\alpha}   \; -  \;
B_{\alpha} \Gamma^{\alpha}  ) \;  \Psi = 0 \; .
\eqno(7.5)
$$

To proceed with an analysis of eq. (7.5), we need one auxiliary relation.
To this end, let us turn again to the used above formula
$ S \Gamma^{a} S^{-1} = \Gamma^{b} L_{b}^{\;\;a} \;$
and take it for an infinitesimal  Lorentz transformation:
$$
(1 + {1 \over 2} \delta \omega_{mn} \;J^{mn}) \; \Gamma^{a} \;
(1 - {1 \over 2} \delta \omega_{kl} \;J^{kl}) =
\Gamma^{b}\;  [ \; \delta^{\;\;a}_{b} + {1 \over 2} \delta \omega_{kl} (V^{kl})_{b}^{\;\;a} \; ] \; ,
$$

\noindent from where it follows
$$
J^{kl} \;\Gamma^{a} - \Gamma^{a} \; J^{kl} = \Gamma^{b} \; (V^{kl})_{b}^{\;\;a} \; .
$$

\noindent Taking into account  the generators $V^{kl}$ explicitly we will have the commutation relation
$$
J^{kl}  \Gamma^{a}  -  \Gamma^{a}  J^{kl}  =   \Gamma^{k}  g^{lk}  -  \Gamma ^{l}  g^{ka} \; .
\eqno(7.6)
$$

\noindent
Now, multiplying eq.  (7.6) by an expression
$ e_{(a)}^{\rho} \; {1 \over 2} e_{(k)}^{\beta} \nabla_{\sigma}  e_{(l)\beta } \; $ we arrive at
$$
\Gamma^{\rho}  B_{\sigma}  - B_{\sigma}  \Gamma^{\rho}  =
\nabla_{\sigma} \Gamma^{\rho}  = \Gamma^{\rho}_{\;\;;\sigma}   \; .
\eqno(7.7)
$$

\noindent Including  (7.7)  in (7.5),  we come to
$$
\bar{\Psi} \stackrel{\leftarrow}{\partial}_{\alpha}   \;  \Gamma^{\alpha}  \Psi  \; +  \;
\bar{\Psi}  \Gamma^{\alpha} \; \stackrel{\rightarrow}{\partial}_{\alpha}
\Psi \; +  \;\bar{\Psi}  \; ( \nabla_{\alpha} \Gamma^{\alpha} )  \;  \Psi = 0 \; ,
$$

\noindent  which may be rewritten  as a generally covariant conserved current law:
$$
\nabla_{\alpha} J^{\alpha} = 0 \; , \;  J^{\alpha} =   \; \bar{\Psi} \Gamma^{\alpha} \Psi    \; .
\eqno(7.8a)
$$

The conserved current $J^{\alpha}(x)$ may be expressed in terms  of generally  covariant tensor
components according to (see (3.20);  a numerical factor is omitted)
$$
J^{\alpha} =   \;  + \sigma \; ( \;  C^{*}  C^{\alpha}   - C  C^{*\alpha }  ) +
(\; \Phi^{*\alpha \beta }  \Phi_{\beta}  -  \Phi^{\alpha \beta } \Phi^{*}_{\beta}\; ) \; .
\eqno(7.8b)
$$

For a while let us concentrate on certain useful and auxiliary formulas and relationships.
In the first place, one can notice that eq. (7.8) may be considered as another form for reading
eq.  (7.5). In other words, the following equality
$$
\nabla_{\alpha}  ( \bar{\Psi} \Gamma^{\alpha}  \Psi  ) =
\bar{\Psi} (\stackrel{\leftarrow}{\partial}_{\alpha} - B_{\alpha})  \;
\Gamma^{\alpha} \Psi  \; +  \; \bar{\Psi}  \Gamma^{\alpha} \;
(\stackrel{\rightarrow}{\partial}_{\alpha}  +   B_{\alpha}) \Psi  \;
\eqno(7.9a)
$$

\noindent holds. With the notation
$$
(\stackrel{\rightarrow}{\nabla}_{\alpha}  +  B_{\alpha} ) \Psi  =
\stackrel{\rightarrow}{D}_{\alpha}  \Psi \;  , \;
\bar{\Psi} (\stackrel{\leftarrow}{\nabla}_{\alpha}  -  B_{\alpha} )   =
\bar{\Psi} \stackrel{\leftarrow}{D}_{\alpha} \;
\eqno(7.9b)
$$

\noindent eq. (7.9a) reads as
$$
\nabla_{\alpha}  ( \bar{\Psi} \Gamma^{\alpha}  \Psi ) =
\bar{\Psi} \stackrel{\leftarrow}{D}_{\alpha} \; \Gamma^{\alpha} \Psi  \; +  \;
\bar{\Psi}  \Gamma^{\alpha} \;\stackrel{\rightarrow}{D}_{\alpha} \; \Psi  \; .
\eqno(7.9c)
$$

It is particularly  remarkable that the formula (7.9c) provides us with a hint about a general
rule for covariant  $\nabla_{\alpha}$ differentiating  any  bilinear combinations of $\Psi$
and  $\bar{\Psi}$. As prescribed by (7.9c), instead of the action of  $\nabla_{\alpha}$
one may act by  $\stackrel{\leftarrow}{D}_{\alpha} $ and by $\stackrel{\rightarrow}{D}_{\alpha} $
from the right and from the left respectively, but not observing  a generally covariant vector characteristic
of the matrix  $\Gamma^{\alpha}(x)$ in the middle.

What is more, such a rule will work always at any complicated bilinear functions.
For instance,  let  us  consider  an  expression
$$
\nabla^{\alpha} [ \; \bar{\Psi} \Gamma^{\rho}(x) \Gamma^{\sigma}(x) \Xi \; ] \; .
$$

\noindent  In  accordance with defining properties of the covariant derivative we can proceed
$$
\nabla^{\alpha} [ \; \bar{\Psi} \Gamma^{\rho} \Gamma^{\sigma} \Xi \; ] =
$$
$$
=   ( \partial_{\alpha} \bar{\Psi} ) \;  \Gamma^{\rho} \Gamma^{\sigma} \; \Xi \;   + \;
\bar{\Psi} \; \Gamma^{\rho}_{\;\; ;\alpha}  \Gamma^{\sigma}  \; \Xi +
 \bar{\Psi} \; \Gamma^{\rho} \Gamma^{\sigma}_{\;\; ; \alpha}  \; \Xi  +
\bar{\Psi} \; \Gamma^{\rho} \Gamma^{\sigma} \; (\partial_{\alpha} \; \Xi  )\; .
$$

\noindent
Hanging  the covariant derivatives of $\Gamma$- matrices with commutator as shown in (7.7), we obtain
$$
\nabla^{\alpha} \; [ \; \bar{\Psi} \Gamma^{\rho} \Gamma^{\sigma} \Xi \; ] =
$$
$$
= ( \partial_{\alpha} \bar{\Psi}) \;  \Gamma^{\rho} \Gamma^{\sigma} \; \Xi  +
\bar{\Psi} \; (\Gamma^{\rho} B_{\alpha}  - B_{\alpha} \Gamma^{\rho}  )
\Gamma^{\sigma} \; \Xi +
\bar{\Psi} \; \Gamma^{\rho} \; (\Gamma^{\sigma} B_{\alpha} - B_{\alpha} \Gamma^{\sigma}  )  \Xi  +
\bar{\Psi} \Gamma^{\rho} \Gamma^{\sigma} \;  ( \partial_{\alpha} \Xi ) \; .
$$

\noindent Two terms  containing $\Gamma^{\rho} B_{\alpha}\Gamma^{\sigma}$   cancel each other,
and the remaining can be rewritten as
$$
\nabla^{\alpha} \; ( \;  \bar{\Psi} \Gamma^{\rho} \Gamma^{\sigma} \Xi \; ) \;
 =  \bar{\Psi} \stackrel{\leftarrow}{D}_{\alpha} \;  \Gamma^{\rho}
 \Gamma^{\sigma} \Xi +
\bar{\Psi}  \Gamma^{\rho} \Gamma^{\sigma}     \;
\stackrel{\rightarrow}{D}_{\alpha} \Xi  \; .
\eqno(7.10a)
$$

\noindent Eq.  (7.10a) is equivalent to
$$
\nabla_{\alpha} (  \Gamma^{\rho} \Gamma^{\sigma} ) =
- B_{\alpha}  \;  \Gamma^{\rho}   \Gamma^{\sigma}  +
\Gamma^{\rho} \Gamma^{\sigma}  \; B_{\alpha}   \; .
\eqno(7.10b)
$$

An analogous formula for a bilinear combination with any tensor structure follows immediately by induction.
Really, let $\Gamma^{(n)}$ be $ \Gamma^{(n)} = \Gamma^{\rho_{1}}  \Gamma^{\rho_{2}} ...
\Gamma^{\rho_{n}}  \;$ ,  then
$$
\nabla_{\alpha}  \Gamma^{(n)}  = \Gamma^{(n)} B_{\alpha} -
B_{\alpha} \Gamma^{(n)} \; ;
\eqno(7.11a)
$$

\noindent    so that for  $\Gamma^{(n+1)} = \Gamma^{(n)} \Gamma^{\rho} $ we can easily derive
$$
\nabla_{\alpha}   \Gamma^{(n+1)}  = \nabla_{\alpha }  [  \Gamma^{(n)} \Gamma^{\rho}  ] =
[\nabla_{\alpha }    \Gamma^{(n)}  ]  \Gamma^{\rho}  +  \Gamma^{(n)} [ \nabla_{\alpha } \Gamma^{\rho} ] =
$$
$$
= [ \Gamma^{(n)} B_{\alpha} - B_{\alpha} \Gamma^{(n)} ]  \Gamma^{\rho}  +
 \Gamma^{(n)} [ \Gamma^{\rho}  B_{\alpha} -B_{\alpha} \Gamma^{\rho} ] =
 \Gamma^{(n)} \Gamma^{\rho} B_{\alpha} - B_{\alpha}  \Gamma^{(n)} \Gamma^{\rho} \; ,
$$

\noindent  which is what we need
$$
\nabla_{\alpha}   \Gamma^{(n+1)}  = - B_{\alpha}  \Gamma^{(n+1)} + \Gamma^{(n+1)}  B_{\alpha}  \; .
\eqno(7.11b)
$$

\noindent  From formulas (7.11) one can produce the commutation relations
$$
\Gamma^{(n) }  \stackrel{\rightarrow}{D_{\sigma}} = \stackrel{\rightarrow}{D_{\sigma}} \Gamma^{(n)} \; , \;\;
\Gamma^{(n)} \stackrel{\leftarrow}{D_{\sigma}} = \stackrel{\leftarrow}{D_{\sigma}} \Gamma^{(n)}\; .
\eqno(7.12)
$$

Now we are ready to write down a Lagrange function that yields eqs.  (7.2)  and  (7.4).
At this, two formulas will be taken into account  (see 7.12)):
$$
\Gamma^{\rho} ( \nabla_{\sigma} + B_{\sigma}) =
( \nabla_{\sigma} + B_{\sigma})  \Gamma^{\rho}   \; , \;\;
\Gamma^{\rho} ( \stackrel{\leftarrow}{\nabla_{\sigma}} - B_{\sigma}) =
( \stackrel{\leftarrow}{\nabla_{\sigma}} - B_{\sigma}) \Gamma^{\rho}   \; , \;\;
\eqno(7.13a)
$$

\noindent  or in a short form
$$
\Gamma^{\rho} \stackrel{\rightarrow}{D_{\sigma}} = \stackrel{\rightarrow}{D_{\sigma}} \Gamma^{\rho} \; , \;\;
\Gamma^{\rho} \stackrel{\leftarrow}{D_{\sigma}} = \stackrel{\leftarrow}{D_{\sigma}} \Gamma^{\rho}\; .
\eqno(7.13b)
$$

\noindent Bearing in mind that the basic equations for $\Psi$ and  $\bar{\Psi}$  look as
$$
[ \; \Gamma^{\alpha} \; \stackrel{\rightarrow}{D}_{\alpha} \; - \; ) \Psi = 0 \; , \;
\eqno(7.14a)
$$
$$
\bar{\Psi}   \; ( \;\Gamma^{\alpha} \;  \stackrel{\leftarrow}{D}_{\alpha}  \; +  m \; ) = 0 \; ,
\eqno(7.14b)
$$

\noindent  a generally covariant Lagrangian for this system can be chosen in the form
$$
L = +  {1 \over 2} \; [ \; \bar{\Psi} \Gamma^{\rho} \stackrel{\rightarrow}{D}_{\rho}  \Psi  -
\bar{\Psi} \Gamma^{\rho}  \stackrel{\leftarrow}{D}_{\rho}  \Psi  \; ] \; -\;  m\; \bar{\Psi} \Psi  \; .
\eqno(7.15a)
$$

\noindent
In terms of tensor components, the $L$ will take on the form
$$
L = {1 \over 2} \;[  \; - \sigma \; ( \; C^{*\alpha} \; \nabla_{\alpha} C  +
C^{\alpha} \; \nabla_{\alpha} C^{*} ) +  \sigma \; (C^{*} \; \nabla_{\alpha} C^{\alpha} +
C \; \nabla_{\alpha} C^{*\alpha} \; ) +
$$
$$
+ (\; \Phi^{*\alpha \beta} \; \nabla_{\alpha} \Phi_{\beta} +
\Phi^{\alpha \beta } \; \nabla_{\alpha} \Phi^{*}_{\beta}\; )
+ ( \;\Phi^{*}_{\alpha} \nabla_{\beta} \Phi^{\alpha \beta} +
\Phi_{ \alpha } \nabla_{\beta} \Phi^{*\alpha \beta }  \;)\; ] -
$$
$$
- m \; [ \; \sigma \; C^{*} C +  C^{*}_{\alpha } C^{\alpha} -
(\Phi^{*}_{\alpha}  C^{\alpha }  +  \Phi_{\alpha } C^{*\alpha} ) +
{1 \over 2} \;\Phi^{*}_{\alpha \beta } \Phi^{\alpha \beta } \; ] \; .
\eqno(7.15b)
$$

The limiting case of a massless particle follows from (7.15) by a formal change
(similar to the flat space-time)
$$
m  \;(\; \bar{\Psi} \;\Psi \; )  \;
\Longrightarrow  \;
\bar{\Psi} P \Psi \; , \;\;
P = \left ( \begin{array}{cccc}
1  &  0  &  0  &  0  \\
0  &  I  &  0  &  0  \\
0  &  0  &  0  &  0  \\
0  &  0  &  0  &  I    \end{array} \right ) \; ,
\eqno(7.16a)
$$

\noindent which transforms the last term in (7.15b)
$$
- m \; [ \; \sigma \; C^{*} C +  C^{*}_{\alpha} C^{\alpha} -
(\Phi^{*}_{\alpha} C^{\alpha} + \Phi_{\alpha} C^{*\alpha} ) +
{1 \over 2} \;\Phi^{*}_{\alpha \beta} \Phi^{\alpha \beta } \; ]
\Longrightarrow
$$
$$
- \; [ \; \sigma \; C ^{*} C +  C^{*}_{\alpha} C^{\alpha}
+ {1 \over 2} \;\Phi^{*}_{\alpha \beta } \Phi^{\alpha \beta} \; ] \; .
\eqno(7.16b)
$$

\subsection*{8.  Massless equation and conformal invariance}

Section 8 deals with a property of conformal invariance as it looks for
15-component theory of massless spin 1 particle.
An original equation is
$$
[ \;  \Gamma ^{\alpha }\; (\partial_{\alpha} \;  +  \;
B_{\alpha } ) \; - P \;  ] \;\Psi    = 0 \; ,
\eqno(8.1)
$$

\noindent projective  matrix  $P$ was defined by  (7.16a). Let the metric tensors of two
curved space-time models differ by a factor - function:
$$
\tilde{g}_{\alpha \beta}(x) = \varphi^{2} (x) \; g_{\alpha \beta} (x) \; .
\eqno(8.2)
$$

\noindent
Correspondingly, we are to use  two vierbeins connected with each other by means  of the same factor
$\varphi$:
$$
\tilde{e}_{(a)}^{\alpha} =  \varphi^{-1}  e_{(a)}^{\alpha}  \; , \;
\tilde{e}_{(a)\alpha} =   \varphi  \;  e_{(a)\alpha}  \; .
\eqno(8.3)
$$

For two sets of Christoffel symbols  we have
$$
\tilde{\Gamma}_{\alpha \beta, \rho}  =  {1 \over 2} \; [\;
\partial_{\alpha} \tilde{g}_{\beta \rho} + \partial_{\beta} \tilde{g}_{\alpha \rho}
- \partial_{\rho} \tilde{g}_{\alpha \beta} \; ] =
$$
$$
= \varphi^{2} \Gamma_{\alpha \beta, \rho} + \varphi \; [
(\partial_{\alpha} \varphi ) g_{\beta \rho} + (\partial_{\beta} \varphi ) g_{\alpha \rho}
- (\partial_{\rho} \varphi ) g_{\alpha \beta}  \; ] \; ,
$$

\noindent or, with the use of the representative
$\tilde{\Gamma}^{\sigma}_{\alpha \beta} = \tilde{g}^{\sigma \rho} \Gamma_{\alpha \beta, \rho}$,
in another form
$$
\tilde{\Gamma}^{\sigma}_{\alpha \beta} = \Gamma^{\sigma}_{\alpha \beta} + { 1 \over \varphi}
[(\partial_{\alpha} \varphi ) \delta^{\sigma}_{\beta}  + (\partial_{\beta} \varphi ) \delta^{\sigma}_{\alpha}
-  g^{\sigma \rho }   (\partial_{\rho} \varphi )  g_{\alpha \beta} \;  ]   \; .
\eqno(8.4)
$$

Let us compare two sets of the Ricci rotation coefficients:
$$
\tilde{\gamma}_{abc}(x) = - [ \; \partial_{\beta} \tilde{e}_{(a)\alpha}   -
\tilde{\Gamma}^{\rho} _{\beta \alpha} \tilde{e}_{(a)\rho} \; ] \;
\tilde{e}_{(b)}^{\alpha} \tilde{e}_{(c)}^{\beta} =
$$
$$
=  -  [ \; \partial_{\alpha}   ( \varphi e_{(a)\alpha} )-
\Gamma^{\rho}_{\beta \alpha} \; \varphi e_{(a)\rho} -
$$
$$
-
{1 \over \varphi }  \;
\left ( \;(\partial_{\beta} \varphi ) \delta^{\rho}_{\alpha}  +
(\partial_{\alpha} \varphi ) \delta^{\rho}_{\beta}
- g^{\rho \sigma}  (\partial_{\sigma} \varphi ) g_{\beta \alpha } \right )\;
\varphi e_{(a)\rho } \;
\; ] \;   {1 \over \varphi^{2} } e_{(b)}^{\alpha} e_{(c)}^{\beta}  \; .
$$

\noindent and further
$$
\tilde{\gamma}_{abc} = -
[ \; \varphi \;  e_{(a)\alpha; \beta}  - {\partial  \varphi \over \partial x ^{\alpha} }
e_{(a)\beta} + e_{(a)}^{\sigma}
{ \partial   \varphi \over \partial x^{\sigma} } g_{\alpha \beta}  \;  ]
 \; {1 \over \varphi^{2} } e_{(b)}^{\alpha} e_{(c)}^{\beta} \; .
$$

\noindent So, the formula we need is
$$
\tilde{\gamma}_{abc} = { 1 \over \varphi } \; \gamma_{abc}  + { 1 \over \varphi^{2}} \;
(\partial_{\sigma} \varphi ) \; [ e_{(b)}^{\sigma}   g_{ac}  - e_{(a)}^{\sigma}  \; g_{bc} ] \;.
\eqno(8.5)
$$

An equation (8.1) in a space-time  with the metric tensor  $\tilde{g}_{\alpha \beta}(x)$ is
$$
[ \; \tilde{\Gamma} ^{\alpha }\; (\partial_{\alpha} \;  +  \;
\tilde{B}_{\alpha } ) \; - P \;  ] \;\tilde{\Psi}    = 0 \; ,
\eqno(8.6)
$$

\noindent or
$$
[\; \Gamma^{c} \; ( \tilde{e}_{(c)}^{\alpha} \partial_{\alpha} +
{1 \over 2} J^{ab} \tilde{\gamma}_{abc} ) - P \; ] \tilde{\Psi} = 0 \; .
\eqno(8.7)
$$

\noindent
Substituting expressions for a new tetrad and new  Ricci coefficients (tilded) in terms of
old (not-tilded) ones, we get to
$$
[ \Gamma^{c} ( e_{(c)}^{\sigma} \partial_{\sigma} + {1 \over 2}  J^{ab} \gamma_{abc} ) +
{ 1 \over 2} \; \Gamma ^{c} J^{ab} \;[ \; ( e_{(b)}^{\sigma}   g_{ac}  -  e_{(a)}^{\sigma}  \; g_{bc} )\;
(\varphi^{-1} \partial_{\sigma}   \varphi)  ] \tilde{\Psi} =  \varphi  P \; \tilde{\Psi} \;.
$$

\noindent and further
$$
[ \;  \Gamma^{\sigma} ( \partial_{\sigma} + B_{\sigma}) +  \; \Gamma _{a} J^{ab} e_{(b)}^{\sigma} \;
   (\varphi^{-1} \partial_{\sigma}  \varphi ) ] \tilde{\Psi} = \varphi  P\; \tilde{\Psi} \;.
\eqno(8.8)
$$

\noindent
With the use of block-representations for all matrices, consider a combination $\Gamma _{a} J^{ab}$:
$$
\Gamma_{a} J^{ab} =
\left ( \begin{array}{cccc}
             0  &  G_{a} V^{ab}  &  0  &  0  \\
             0  &  0  &  0  &  K_{a} (V\otimes V)^{ab} \\
             0  &  0  &  0  &  K_{a} (V\otimes V)^{ab} \\
             0  &  0  &  \Lambda_{a} V^{ab} &  0
\end{array} \right ) \; .
\eqno(8.9)
$$

\noindent
Allowing for relations
$$
(G_{a})_{(0)}^{\;\;\;l} (V^{ab})_{l}^{\;\;k} = + 3 g^{bk} \; , \; \;
(\Lambda_{a})_{ps}^{\;\;\;\;l} (V^{ab})_{l}^{\;\;k} =  \delta_{ps}^{bk} \; ,
$$
$$
(K_{a})_{l}^{\;\;mn} [ (V \otimes V)^{ab}]_{mn}^{\;\;\;\;ps} =
2 (- g^{bp} \delta_{l}^{s} + g^{bs} \delta_{l}^{p} ) \; ,
$$

\noindent an expression for $\Gamma_{a} J^{ab}$ is  led  to
$$
\Gamma_{a} J^{ab} =
\left ( \begin{array}{cccc}
          0  &  + 3 g^{bk}  &     0             &                             0                         \\
          0  &      0       &     0             &  2 (-g^{bp} \delta_{l}^{s} + g^{bs} \delta_{l}^{p} )  \\
          0  &      0       &     0             &  2 (-g^{bp} \delta_{l}^{s} + g^{bs} \delta_{l}^{p} )  \\
          0  &      0       & \delta_{ps}^{bk}  &                             0
\end{array} \right ) \; .
\eqno(8.10)
$$

\noindent From this, with taking in mind eq. (2.12), we find
$$
\Gamma_{a} J^{ab} =
\left ( \begin{array}{cccc}
        0  &   3 (G^{b})_{(0)}^{\;\;k} &             0                &          0             \\
        0  &           0               &             0                &  2(K^{b})_{l}^{\;\;ps} \\
        0  &           0               &             0                &  2(K^{b})_{l}^{\;\;ps} \\
        0  &           0               &  (\Lambda^{b})_{ps}^{\;\;k}  &          0
\end{array} \right ) =
\Gamma^{b} \left ( \begin{array}{cccc}
0  &  0  &  0  &  0     \\
0  &  +3 I  &  0  &  0  \\
0  &  0  &  +I  &  0    \\
0  &  0  &  0  &  +2 I
\end{array} \right ) \; .
\eqno(8.11a)
$$

\noindent In the end, we  arrive at
$$
\Gamma_{a} J^{ab} =
\Gamma^{b} (+ 3 P'_{4}  + P_{4} + 2 P_{6}) \; ,
\eqno(8.11b)
$$

\noindent where a special notation for projective operators on 4-dimensional and 6-dimensional
subspaces is used. Hence, eq. (8.9) reads as
$$
\Gamma^{\sigma} [ \;  \partial_{\sigma} + B_{\sigma} +  (3 P'_{4}  + P_{4} + 2 P_{6})
 ( \varphi^{-1} \partial_{\sigma} \varphi )\; ] \; \tilde{\Psi} =   \varphi P   \tilde{\Psi}  \; .
\eqno(8.12)
$$

\noindent With the use of block-forms of all matrices involved in (8.12) and  taking the notation
$$
\tilde{\Psi} =  ( \tilde{\Phi} , \tilde{C} , \tilde{A} , \tilde{F} )  \; ,
\eqno(8.13)
$$

\noindent eq.  (8.12) is shown to be equivalent to a set (compare with  (6.21))
$$
G^{\alpha} [ \partial_{\alpha} + V_{\alpha}  +
3 \partial_{\alpha} \ln \varphi  ]  \; \tilde{C} = \varphi \tilde{\Phi} \; ,
\eqno(8.14a)
$$
$$
K^{\alpha} [ \partial_{\alpha} + (V \otimes V)_{\alpha}  + 2
\partial_{\alpha} \ln \varphi  ]  \; \tilde{F}   =  \varphi \tilde{C} \; ,
\eqno(8.14b)
$$
$$
\sigma \Delta^{\alpha} \partial_{\alpha}  \tilde{\Phi}  +  K^{\alpha} [ \partial_{\alpha} +
 (V \otimes V)_{\alpha}  + 2 \partial_{\alpha} \ln \varphi  ]  \; \tilde{F}   = 0 \; ,
\eqno(8.14c)
$$
$$
\Lambda^{\alpha}  [ \partial_{\alpha} + V_{\alpha} +
\partial_{\alpha} \ln \varphi  ]  \; \tilde{A} = \varphi \tilde{F}
\eqno(8.14d)
$$

\noindent
Let us find out how will look these equations after substitution the field components in the form
$$
\tilde{A}_{l} = \varphi^{-1} A_{l} \; , \;  \tilde{F}_{mn} =  \varphi^{-2} F_{mn} \; ,
\tilde{C}_{l} = \varphi^{-3} C_{l} \; , \;  \tilde{\Phi}   =  \varphi^{-4} \Phi   \; .
\eqno(8.15)
$$

\noindent It is straightforward to obtain
$$
G^{\alpha} [ \partial_{\alpha} + V_{\alpha}] C = \Phi \; ,
\eqno(8.16a)
$$
$$
K^{\alpha} [ \partial_{\alpha} + (V \otimes V)_{\alpha} ]\; F = C \; ,
\eqno(8.16b)
$$
$$
\varphi^{2} \; \sigma \Delta^{\alpha} \partial_{\alpha}  \tilde{\Phi}  +
K^{\alpha} [ \partial_{\alpha} + (V \otimes V)_{\alpha} ] \; F   = 0 \; ,
\eqno(8.16c)
$$
$$
\Lambda^{\alpha} [ \partial_{\alpha} + V_{\alpha} ] \; A = F \; .
\eqno(8.16d)
$$

Take notice that apparently eq.  (8.16c) does not coincides with a respective equation in
$g_{\alpha \beta}$-space. However, if the vanishing of supplementary scalar and vector components
$\tilde{\Phi}  = 0 \; , \;\; \tilde{C}  = 0 \; $ is taken into consideration, then effectively remaining two
equations in (8.16) will read as
$$
K^{\alpha} [ \partial_{\alpha} +
(V \otimes V)_{\alpha}   ]  \; F   = 0 \; ,
$$
$$
\Lambda^{\alpha} [ \partial_{\alpha} + V_{\alpha}(x) ] \;  A = F \; ,
\eqno(8.17)
$$

\noindent which are  precisely  required ones  for a set $(\Phi  = 0 , C = 0  , A , F )$ in
$g_{\alpha \beta}$-space. This means that the property of conformal invariance holds.

To avoid misleading it should be mentioned that  just accomplished  study of conformal invariance
in 15-component theory of a massless vector particle does not add much to yet known for a conventional
10-component theory; rather it lies in the freedom from any internal conflict in the  theory.
This is what should be expected because  of equivalence demonstrated  above between 15- and 10-component
models in  a free particle  case.

\subsection*{9. Canonical energy-momentum tensor}

In  Sec. 9 we are going to study a question about energy-momentum tensor for a generalized spin 1 particle.
At this it is convenient  to exploit a matrix formalism, when  two basic equations are given by (for much
generality, an external electromagnetic field will be taken into account)
$$
( \; \Gamma^{\alpha } \; \stackrel{\rightarrow}{D}_{\alpha}   \; - \; m \;  )  \;\Psi   = 0 \; , \;\;
\bar{\Psi}  \;  (  \; \Gamma^{\alpha}  \stackrel{\leftarrow}{D}_{\alpha} \; + \; m  \; ) = 0 \; ,
\eqno(9.1a)
$$

\noindent where
$$
\stackrel{\rightarrow}{D}_{\alpha} =  \stackrel{\rightarrow}{\nabla}_{\alpha} + B_{\alpha}  - i g A_{\alpha} \; , \;\;
\stackrel{\leftarrow}{D}_{\alpha} =  \stackrel{\leftarrow}{\nabla}_{\alpha} - B_{\alpha} + i g  A_{\alpha} \; ,
\;\; g \equiv e /\hbar c  \; .
\eqno(9.1b)
$$

Let a tensor quantity $W^{\;\; \alpha}_{\beta}$ be
$$
W^{\;\; \alpha}_{\beta} =  \bar{\Psi} \Gamma^{\alpha}
\stackrel{\rightarrow}{D}_{\beta} \Psi =  \bar{\Psi} \Gamma^{\alpha}
( \stackrel{\rightarrow}{\partial}_{\beta} + B_{\beta})  \Psi  \;
- \; i g   A_{\beta} \; (\bar{\Psi} \Gamma^{\alpha}  \Psi  )\; .
\eqno(9.2)
$$

\noindent  The aim is to calculate a divergence over it. To this end, act on the first equation in (9.1)
from the left by an    operator $\bar{\Psi} \stackrel{\rightarrow}{D}_{\beta}$, and multiply the second
from  the  right  by  на   $\stackrel{\rightarrow}{D}_{\beta} \Psi $,   and sum  results, then we  have
$$
\bar{\Psi} \stackrel{\rightarrow}{D}_{\beta} \;
  \Gamma^{\alpha}  \stackrel{\rightarrow}{D}_{\alpha} \Psi  \; + \;
\bar{\Psi}  \stackrel{\leftarrow}{D}_{\alpha} \Gamma^{\alpha}  \;
\stackrel{\rightarrow}{D}_{\beta} \Psi  = 0  \; ,
\eqno(9.3a)
$$

\noindent an further
$$
\bar{\Psi} \Gamma^{\alpha} \; [ \; (\stackrel{\rightarrow}{D}_{\beta}
\stackrel{\rightarrow}{D}_{\alpha} - \stackrel{\rightarrow}{D}_{\alpha}
\stackrel{\rightarrow}{D}_{\beta}) \; + \;
\stackrel{\rightarrow}{D}_{\alpha} \stackrel{\rightarrow}{D}_{\beta}\;  ]\;
\Psi  \; +
\;\bar{\Psi} \stackrel{\leftarrow}{D}_{\alpha}  \Gamma^{\alpha} \;
\stackrel{\rightarrow}{D}_{\beta} \Psi  = 0 \; .
\eqno(9.3b)
$$

\noindent Now, placing a term with commutator
$[\stackrel{\rightarrow}{D}_{\beta}, \stackrel{\rightarrow}{D}_{\alpha}]_{-}$ on the right, we get to
$$
\bar{\Psi} (\;   \Gamma^{\alpha} \stackrel{\rightarrow}{D}_{\alpha}  +
 \stackrel{\leftarrow}{D}_{\alpha}   \Gamma^{\alpha}\; ) \;
\stackrel{\rightarrow}{D}_{\beta}   \Psi   = \bar{\Psi}  \Gamma^{\alpha}
[ \stackrel{\rightarrow}{D}_{\alpha} , \stackrel{\rightarrow}{D}_{\beta} ]_{-} \Psi \; .
\eqno(9.3c)
$$

\noindent Let us show that an expression on the left can be thought as a pure divergence of
$W^{\;\; \alpha}_{\beta}$. We are to start with
$$
\bar{\Psi} (\;   \Gamma^{\alpha} \stackrel{\rightarrow}{D}_{\alpha}  +
 \stackrel{\leftarrow}{D}_{\alpha}   \Gamma^{\alpha}\; ) \; \stackrel{\rightarrow}{D}_{\beta}   \Psi   =
$$
$$
=
\bar{\Psi} [ \; \Gamma^{\alpha} (\stackrel{\rightarrow}{\nabla}_{\alpha} + B_{\alpha} - i g  A_{\alpha})
 + ( \stackrel{\leftarrow}{\nabla}_{\alpha} - B_{\alpha} + i g A_{\alpha})
\Gamma^{\alpha} \; ] \; \stackrel{\rightarrow}{D}_{\beta} \Psi =
$$
$$
= \bar{\Psi} \; [ \; \Gamma^{\alpha} \stackrel{\rightarrow}{\nabla}_{\alpha}
+  ( \Gamma^{\alpha}  B_{\alpha}  -  B_{\alpha} \Gamma^{\alpha}  )  +
\stackrel{\leftarrow}{\nabla}_{\alpha} \Gamma^{\alpha} \; ] \;  \stackrel{\rightarrow}{D}_{\beta} \Psi   \; .
\eqno(9.4a)
$$

\noindent Taking into consideration eq.  (7.7),  relation (9.4a) can be transformed  into
$$
\bar{\Psi} \; ( \;\Gamma^{\alpha}  \stackrel{\rightarrow}{D}_{\alpha}
+ \stackrel{\leftarrow}{D}_{\alpha}   \Gamma^{\alpha}
\; ) \; \stackrel{\rightarrow}{D}_{\beta}   \Psi   =
$$
$$
=  \bar{\Psi} (\; \stackrel{\leftarrow}{\nabla}_{\alpha}
 \Gamma^{\alpha} \; +   \Gamma^{\alpha}_{\;\; ; \alpha} + \Gamma^{\alpha}
  \stackrel{\rightarrow}{\nabla}_{\alpha} \; ) \;  \stackrel{\rightarrow}{D}_{\beta} \Psi  =
 \nabla_{\alpha}  ( \bar{\Psi} \Gamma^{\alpha} \stackrel{\rightarrow}{D}_{\beta} \Psi  )  \; .
\eqno(9.4b)
$$

\noindent
Hence, eq.  (9.3c) is  equivalent to
$$
\nabla_{\alpha}\; [\; W^{\;\;\;\alpha}_{\beta}\;  ] =
\bar{\Psi} \Gamma^{\alpha}  [ \stackrel{\rightarrow}{D}_{\alpha} ,
\stackrel{\rightarrow}{D}_{\beta} ]_{-} \Psi \; .
\eqno(9.5)
$$

\noindent
Let us consider in more detail the commutator
$$
[\stackrel{\rightarrow}{D}_{\alpha}, \stackrel{\rightarrow}{D}_{\beta} ]_{-} =
- i g  F_{\alpha \beta} \; + \; D_{\alpha \beta} \; ,
\eqno(9.6a)
$$

\noindent where
$$
F_{\alpha \beta} = {\partial A_{\beta} \over \partial x^{\alpha }}  -
                   {\partial A_{\alpha}\over  \partial x^{\beta  }} \; , \;\;
D_{\alpha \beta} =   {\partial B_{\beta } \over \partial  x^{\alpha} }      -
                     {\partial B_{\alpha} \over \partial  x^{\beta } }
+  ( B_{\alpha} B_{\beta} - B_{\beta} B_{\alpha}  ) \; .
\eqno(9.6b)
$$

\noindent
For first term on  the right in expression for $D_{\alpha \beta}$,  it is easily to  obtain a representation
$$
\partial_{\alpha } B_{\beta}  -
\partial_{\beta }  B_{\alpha} = \nabla_{\alpha} B_{\beta} -  \nabla_{\beta} B_{\alpha} =
$$
$$
= {1 \over 2} J^{ab} \nabla_{\alpha} \; ( \; e_{(a)}^{\nu} e_{(b)\nu ; \beta} \;) \; -   \;
  {1 \over 2} J^{ab} \nabla_{\beta}  \; ( \;e_{(a)}^{\nu} e_{(b)\nu ; \alpha} \; ) =
$$
$$
= {1 \over 2} J^{ab} e_{(a)}^{\nu} \; [\; e_{(b)\nu ; \beta ; \alpha } -
e_{(b)\nu ; \alpha ; \beta }\; ] \; +
{1 \over 2 } J^{ab} \; [\; e_{(a) \nu ; \alpha } e_{(b) ; \beta}^{\nu} -
e_{(a) \nu ; \beta} e_{(b) ; \alpha}^{\nu} \; ] \; .
\eqno(9.7)
$$

\noindent
For second one it follows
$$
( B_{\alpha} B_{\beta} - B_{\beta} B_{\alpha}  ) =
$$
$$
=
(\; {1 \over 2} J^{ab} e_{(a)}^{\nu} e_{(b)\nu ; \alpha }  ) \;
(\; {1 \over 2} J^{kl} e_{(k)}^{\mu} e_{(l)\mu ; \beta } \; ) \; - \;
(\; {1 \over 2} J^{kl} e_{(k)}^{\mu} e_{(l)\mu ; \beta }  ) \;
(\; {1 \over 2} J^{ab} e_{(a)}^{\nu} e_{(b)\nu ; \alpha } \; ) \; =
$$
$$
= {1 \over 4} \left ( J^{ab} J^{kl}    -  J^{kl}  J^{ab} \right  ) \left [
\;   (  e_{(a)}^{\nu} e_{(b)\nu ; \alpha }  ) \;
\; e_{(k)}^{\mu} e_{(l)\mu ; \beta } \; )  \right ] \; .
$$

\noindent Now, with the use of commutation relations (see (6.11))
$$
[J^{ab},J^{kl}]_{-} = ( -J^{kb} g^{la} + J^{lb} g^{ka} ) - (  - J^{ka} g^{lb}  + J^{la} g^{kb} ) \; ,
$$

\noindent we get
$$
( B_{\alpha} B_{\beta} - B_{\beta} B_{\alpha}  ) =
 - {1 \over 2 } J^{ab} \; [\; e_{(a) \nu ; \alpha } e_{(b) ; \beta}^{\nu} -
                      e_{(a) \nu ; \beta} e_{(b); \alpha}^{\nu} \; ] \; .
\eqno(9.8)
$$

\noindent Accounting for eqs.  (9.7) and  (9.8), expression for  $D_{\alpha \beta}$
can be led to the  form
$$
D_{\alpha \beta} =  {1 \over 2} J^{ab}\; e_{(a)}^{\nu} \;
[\; e_{(b)\nu ; \beta ; \alpha } - e_{(b)\nu ; \alpha ; \beta }\; ] =
$$
$$
=
{1 \over 2} J^{ab}\; e_{(a)}^{\nu} \;  [ \;e_{(b)} ^{\rho} R_{\rho \nu \beta \alpha} (x)\; ] =
{1 \over 2 } J^{\nu \rho} (x)\; R_{\nu \rho \alpha \beta }(x) \; ,
\eqno(9.9)
$$

\noindent where  $R_{\nu \rho \alpha \beta }(x) $ designates  a Riemann curvature tensor.
Thus, eq.  (9.5) reads as
$$
\nabla_{\alpha} W^{\;\;\;\alpha}_{\beta}  = - i g \; J^{\alpha}    F_{\alpha \beta }          \; + \;
{1 \over 2} \; R_{\nu \rho \alpha \beta}  \; \bar{\Psi} \Gamma^{\alpha}  J^{\nu \rho} \Psi  \; .
\eqno(9.10)
$$

\noindent
It is useful to perform some transformation over second term on the right:
$$
{1 \over 2}  \;  R_{\nu \rho \alpha \beta}  \;
\bar{\Psi} \; \Gamma^{\alpha} J^{\nu \rho} \; \Psi  =
$$
$$
=
{1 \over 2} \; R_{\nu \rho \alpha \beta}  \; \bar{\Psi} \;
\left ( {1 \over 2} [\;
\Gamma^{\alpha} J^{\nu \rho} -
J^{\nu \rho}  \Gamma^{\alpha}\;  ]  +
{1 \over 2}  [\; \Gamma^{\alpha} J^{\nu \rho} +
J^{\nu \rho} \Gamma^{\alpha}  \; ] \right ) \; \Psi \; .
$$

\noindent Further, with the use of commutation relation (see (7.6))
$$
\Gamma^{\alpha}  J^{\nu \rho}   -   J^{\nu \rho}   \Gamma^{\alpha}    =
g^{\alpha \nu}(x) \Gamma^{\rho} - g^{\alpha \rho} (x) \Gamma^{\nu} \; ,
$$

\noindent one can produce
$$
{1 \over 2} \; R_{\nu \rho \alpha \beta}  \; \bar{\Psi} \; \Gamma^{\alpha} J^{\nu \rho} \; \Psi  =
$$
$$
=
{1 \over2 } R_{\alpha \beta} J^{\alpha}  + {1 \over 4} \; R_{\nu \rho \alpha \beta}  \; \bar{\Psi} \;
(\Gamma^{\alpha}  J^{\nu \rho} \; + \; J^{\nu \rho} \Gamma^{\alpha}  ) \; \Psi \; .
\eqno(9.11)
$$

\noindent
Correspondingly, eq. (9.10) takes on the form
$$
\nabla_{\alpha} \; [ \; W^{\;\;\;\alpha}_{\beta} \; ] = -i g \;J^{\alpha}   F_{\alpha \beta } \; + \;
{1 \over2 } \;  J^{\alpha} R_{\alpha \beta} + {1 \over 4} \; R_{\nu \rho \alpha \beta}  \; \bar{\Psi}
(\Gamma^{\alpha}  J^{\nu \rho} + J^{\nu \rho} \Gamma^{\alpha}  )\Psi \; .
\eqno(9.12a)
$$

\noindent This  relation can be rewritten as
$$
\nabla_{\alpha} \; [ \; W^{\;\;\;\alpha}_{\beta} \; ]  =
J^{\alpha}  ( \;-i g \; F_{\alpha \beta }\; + \;
{1 \over2 } R_{\alpha \beta} \; ) + {1 \over 4}
\; R_{\nu \rho \alpha \beta}  \; \bar{\Psi}
(\Gamma^{\alpha}  J^{\nu \rho} + J^{\nu \rho} \Gamma^{\alpha}  ) \; \Psi \; ,
\eqno(9.12b)
$$

\noindent which generalizes a known formula  established by V.A. Fock [49, eq. (56)] at studying
a spin 1/2  particle on the background of a curved space-time model.
A single formal difference consists in occurrence of one  additional term proportional to
Riemann curvature tensor on the left.

It may be checked quite easily that, in  the established formula  (9.12b) being applied to spin 1/2
particle case, this additional $R$-dependent term will vanish identically.
To this end, let us consider  more closely a combination of Dirac matrices
$$
(\gamma^{\alpha}  j^{\nu \rho} + j^{\nu \rho} \gamma^{\alpha}  ) =
{1 \over 4} [\; \gamma^{\alpha} (\gamma^{\nu} \gamma^{\rho} - \gamma^{\rho} \gamma^{\nu}) +
(\gamma^{\nu} \gamma^{\rho} - \gamma^{\rho} \gamma^{\nu}) \; \gamma^{\alpha}  \; ]
$$

\noindent Multiplying the known formula for the  product  of the ordinary  Dirac matrices [12]
$$
\gamma^{a} \gamma^{b} \gamma^{c} =
\gamma^{a} g^{bc} -  \gamma^{b} g^{ac} + \gamma^{c} g^{ab} + i \gamma^{5} \epsilon^{abcd} \gamma_{d}
\eqno(9.13a)
$$

\noindent by the following tetrad construction $e^{\alpha}_{(a)} e^{\beta}_{(b)} e^{\rho}_{(c)}$,
one obtains
$$
\gamma ^{\alpha }(x) \; \gamma ^{\beta }(x) \; \gamma ^{\rho }(x)  =
[\;  \gamma ^{\alpha }(x) \; g^{\beta \rho }(x) \;  -  \; \gamma ^{\beta }(x) \; g^{\alpha \rho }(x) \; +
$$
$$
+ \; \gamma ^{\rho }(x) \; g^{\alpha \beta }(x) \; + \; i \gamma ^{5} \;
\epsilon ^{\alpha \beta \rho \sigma }(x)\; \gamma _{\sigma }(x)
\; ] \;  .
\eqno(9.13b)
$$

\noindent Here a generally covariant Levi-Civita symbol is given by
$$
\epsilon ^{\alpha \beta \rho \sigma }(x) =
e^{\alpha}_{(a)} e^{\beta}_{(b)} e^{\rho}_{(c)} e^{\sigma}_{(d)} \epsilon^{abcd}  \; .
\eqno(9.13c)
$$

\noindent Now, with the use of (9.13b), one can produce
$$
\gamma^{\alpha} (x) j^{\nu \rho} (x) + j^{\nu \rho}(x) \gamma^{\alpha} (x) =
i \gamma^{5}\; \epsilon^{\alpha \nu \rho \sigma}(x) \; \gamma_{\sigma}(x) \; .
\eqno(9.14a)
$$

\noindent So, we have arrived to
$$
{1 \over 4}\; R_{\nu \rho \alpha \beta} (x) \; \bar{\Psi}
(\Gamma^{\alpha}  J^{\nu \rho} + J^{\nu \rho} \Gamma^{\alpha}  )\Psi =
{1 \over  4}  \; R_{\nu \rho \alpha \beta}(x)  \;
 \epsilon^{\alpha \nu \rho \sigma}(x) \;
\bar{\Psi}    \gamma^{5}  \; \gamma_{\sigma}(x)  \Psi  \equiv 0 \; ;
\eqno(9.14б)
$$

\noindent where it is taken into account the known symmetry of the curvature tensor under cyclic permutation
over  any three indices so that convolution of $R^{...}$ with $\epsilon_{....}$ over  three indices equals to zero.

Returning again to (9.12b), with the use of the properties of invariant-form matrix  $\eta$
(see  (3.3) and (7.1b)):
$$
 \eta^{-1} [\Gamma^{\alpha}(x)]^{+}  \eta = - \Gamma^{\alpha}(x) \; , \;
\eta^{-1} [J^{\nu \rho}(x)]^{+} \eta = - J^{\nu \rho }(x)  \; ,
\eqno(9.15)
$$

\noindent one can find that on the right in (9.12a)
the first and the third terms are real-valued, whereas the second is  imaginary:
$$
\nabla_{\alpha} W^{\;\;\;\alpha}_{\beta}  =  Re (x) +  i \; Im (x) \;, \;\;
Im (x) = -   \; {i \over2 }   J^{\alpha}  R_{\alpha \beta}  \;  ,
$$
$$
Re (x) =  \;- i g \; J^{\alpha}   F_{\alpha \beta }      \; + \; {1 \over 4 }
\; R_{\nu \rho \alpha \beta}  \; \bar{\Psi} (\Gamma^{\alpha}  J^{\nu \rho} +
J^{\nu \rho} \beta^{\alpha}  ) \Psi \; .
\eqno(9.16)
$$

\noindent
Now, we need to isolate real and imaginary parts on the left in (9.12a) too.
To this end, let us find a complex conjugate tensor  $(W^{\alpha} _{\;\;\beta})^{*}$:
$$
(W^{\;\;\;\alpha}_{\beta})^{+} = [ \; \Psi^{+} \eta \Gamma^{\alpha}
(\nabla_{\beta} + B_{\beta} - ig A_{\beta}) \Psi \; ]^{+} =
$$
$$
=
- \; \bar{\Psi} \Gamma^{\alpha}
(\stackrel{\leftarrow}{\nabla}_{\beta} - B_{\beta} + ig A_{\beta}) \Psi =
- \; \bar{\Psi} \stackrel{\leftarrow}{D}_{\beta} \Gamma^{\alpha}  \Psi \; .
\eqno(9.17)
$$

\noindent With the notation
$$
Re \;( W^{\;\;\; \alpha}_{\beta} ) = {1 \over 2} \left
  [ \; W^{\;\;\;\alpha}_{\beta} +
    ( W^{\;\;\; \alpha}_{\beta} )^{+} \; \right ] = T^{\;\;\;\alpha}_{\beta} \; ,
$$
$$
Im  \; ( W^{\;\;\; \alpha}_{\beta} ) = {1 \over 2 i } \left
   [ \; W^{\;\;\; \alpha}_{\beta} -
     (  W^{\;\;\; \alpha}_{\beta} )^{+} \; \right ] =
U^{\;\;\; \alpha}_{\beta}\; ,
\eqno(9.18)
$$

\noindent eq.  (9.16) will split into two  real-valued ones
$$
\nabla_{\alpha} (T^{\;\;\; \alpha}_{\beta}) =
- i g \; J^{\alpha}(x)  F_{\alpha \beta}    \; + \; {1 \over 4 }
\; R_{\nu \rho \alpha \beta}  \; \bar{\Psi} \;
(\Gamma^{\alpha}  J^{\nu \rho} + J^{\nu \rho} \Gamma^{\alpha}  ) \; \Psi  \; .
\eqno(9.19)
$$
$$
\nabla_{\alpha} (U^{\;\;\; \alpha}_{\beta}) = -{i \over2 }  J^{\alpha } R_{\alpha \beta}  \; ,
\eqno(9.20)
$$

As it is readily checked,  eq. (9.20) represent in essence a direct consequence of the
conserved current law. Really, in accordance with definition for $U^{\;\;\; \alpha}_{\beta}$ we have
$$
U^{\;\;\; \alpha}_{\beta} = {1 \over 2i} \left [
\bar{\Psi} \Gamma^{\alpha} ( \partial_{\beta} + B_{\beta} - i A_{\beta} ) \Psi +
\bar{\Psi} ( \stackrel{\leftarrow}{\partial}_{\beta} - B_{\beta} + i A_{\beta}) \Gamma^{\alpha} \Psi
\right ] =
$$
$$
= {1 \over 2 i } \nabla_{\beta} \; ( \bar{\Psi} \Gamma^{\alpha} \Psi ) =
{1 \over 2i} \nabla_{\beta} J^{\alpha}\; .
\eqno(9.21)
$$

\noindent
Therefore, eq. (9.19) can be led to the form
$$
\nabla_{\alpha} \nabla_{\beta} J^{\alpha} = R_{\alpha \beta} J^{\alpha}  \; ,
\eqno(9.22a)
$$

\noindent and further
$$
(\nabla_{\alpha} \nabla_{\beta}    - \nabla_{\beta} \nabla_{\alpha}) J^{\alpha} \; + \;
\nabla_{\beta} \nabla_{\alpha} J^{\alpha}  =  J^{\alpha} \; R_{\alpha \beta}   \; .
$$

\noindent From this, with bearing in mind the conservation law for  $J^{\alpha}$, it follows an identity
$$
J^{\rho} (x) \;  R_{\rho\;\;\;    \beta \alpha }^{\;\;\alpha}(x)  \equiv
 J^{\alpha}(x)  \;  R_{\alpha \beta} (x)   \; .
\eqno(9.22b)
$$

\noindent Thus,  eq.  (9.20)  does not contain anything new in addition to current conservation law.
As for eq.  (9.19), for  $T^{\;\;\; \alpha}_{\beta}$ we have
$$
T^{\;\;\;\alpha}_{\beta} =
{1 \over 2} [ \bar{\Psi} \Gamma^{\alpha} \stackrel{\rightarrow}{D}_{\beta} \Psi -
\bar{\Psi} \Gamma^{\alpha} \stackrel{\leftarrow}{D}_{\beta} \Psi ] =
$$
$$
= {1 \over 2 } [ \bar{\Psi} \Gamma^{\alpha} (\stackrel{\rightarrow}{\nabla}_{\beta} + B_{\beta}) \Psi -
\bar{\Psi} \Gamma^{\alpha} (\stackrel{\leftarrow}{\nabla}_{\beta} - B_{\beta}) \Psi ]  -
ig J^{\alpha} A_{\beta} \; .
\eqno(9.23a)
$$

\noindent and a conservation law reads as follows
$$
\nabla_{\alpha} (T^{\;\;\; \alpha}_{\beta}) =
- i g \; J^{\alpha}  F_{\alpha \beta}    \; + \;
{1 \over 4 } \; R_{\nu \rho \sigma \beta}  \; \bar{\Psi} \;
(\Gamma^{\sigma}  J^{\nu \rho} + J^{\nu \rho} \Gamma^{\sigma}  ) \; \Psi  \; .
\eqno(9.23b)
$$

To proceed with eq. (9.23b), it is a time to remember some facts about ambiguity in determining
any  energy-momentum tensor. In the Minkowski space-time such freedom in  its determining
is described as follows: if $ T_{b}^{\;\;a}(x)$  obeys an equation
$$
\partial_{a} T_{b}^{\;\;a} = 0 \; ,
$$

\noindent then another tensor
$$
\bar{T}_{b}^{\;\;a}(x) = T_{b}^{\;\;a}(x) + \partial_{c} \; [ \;  \Omega_{b}^{\;\;[ac]}(x) \; ] \; , \;
\; \; where \;\; \Omega_{b}^{\;\;[ac]}(x) = - \Omega_{b}^{\;\;[ca]}(x) \;
\eqno(9.24a)
$$

\noindent satisfies the same equation as well
$$
 \partial_{a} \bar{T}_{b}^{\;\;a} = 0 \;
$$

\noindent Obviously, simultaneous existence of the two  conservation laws  is insured by
an elementary  formula
$$
\partial_{a} \partial_{c} \;  \Omega_{b}^{\;\;[ca]} (x) \equiv  0 \; .
\eqno(9.24b)
$$

As for a curved space-time model, such an equivalence between tensors $T_{\beta}^{\;\; \alpha}(x) $
and  $\bar{T}_{\beta}^{\;\; \alpha }(x)$  holds as well but  in a more complicated manner.
Indeed, let two tensors be related to each other by
$$
\bar{T}_{\beta}^{\;\; \alpha } (x) = T_{\beta}^{\;\; \alpha }(x) +
\nabla_{\rho} \; [ \; \Omega_{\beta}^{\;\; [\alpha \rho]} (x) \; ] \; .
\eqno(9.25)
$$

\noindent Acting on both sides  by operation of covariant differentiation $\nabla_{\alpha}$, one  produces
$$
\nabla_{\alpha}  \bar{T}_{\beta}^{\;\; \alpha } (x) =
\nabla_{\alpha}  T_{\beta}^{\;\; \alpha } (x)  +
\nabla_{\alpha } \; [ \; \nabla_{\rho} \Omega_{\beta}^{\;\; [\alpha \rho]} (x) \; ]\; .
\eqno(9.26)
$$

\noindent Now, bearing in  mind symmetry properties of the curvature tensor, one obtains
$$
\nabla_{\alpha} \; [ \;\nabla_{\rho} \Omega_{\beta}^{\;\; [\alpha \rho]} (x)\; ] =
{1 \over 2} [ R_{\alpha \rho \beta}^{\;\;\;\;\;\;\sigma}  \; \Omega_{\sigma}^{\;\;[\alpha \rho]} +
R_{\alpha \rho\;\;\;\sigma}^{\;\;\;\;\alpha} \; \Omega_{\beta}^{\;\;[\sigma \rho]} +
R_{\alpha \rho\;\;\;\sigma}^{\;\;\;\;\rho} \; \Omega_{\beta}^{\;\;[\alpha \sigma]} ] =
$$
$$
= {1 \over 2} [ R^{\beta \sigma}_{\;\;\;\;\; \alpha \rho} \;
\Omega_{\sigma}^{\;\;[\alpha \rho]} -
 R_{\rho \sigma} \; \Omega_{\beta}^{\;\;[\sigma \rho]} -
R_{\alpha \sigma} \;  \Omega_{\beta}^{\;\;  [\alpha \sigma]} ] \; ,
$$

\noindent thus
$$
\nabla_{\alpha} \; [ \;\nabla_{\rho} \Omega_{\beta}^{\;\; [\alpha \rho]} (x)\; ] =
{1 \over 2} R_{\beta\sigma \nu \rho} \Omega^{\sigma[\nu \rho]} \; .
\eqno(9.27)
$$

\noindent In the end, eq. (9.26) reads as
$$
\nabla_{\alpha}  \bar{T}_{\beta}^{\;\;\; \alpha} (x) =
\nabla_{\alpha}  T_{\beta}^{\;\;\;  \alpha } (x)  +
{1 \over 2} R_{\beta\sigma \nu \rho} (x) \; \Omega^{\sigma[\nu \rho]}(x) \; .
\eqno(9.28)
$$

\noindent which, on  accounting eq.(9.23b),  takes the form
$$
\nabla_{\alpha}  \bar{T}_{\beta}^{\;\; \alpha } (x) =
- i g \; J^{\alpha}  F_{\alpha \beta}    \; + \;
{1 \over 4 } \; R_{\nu \rho \sigma \beta}  \; \bar{\Psi} \;
(\Gamma^{\sigma}  J^{\nu \rho} + J^{\nu \rho} \Gamma^{\sigma}  ) \; \Psi  \; +
$$
$$
 + \;  {1 \over } R_{\beta \sigma \nu \rho} (s) \Omega^{\sigma[\nu \rho]}(x) \; .
\eqno(9.29a)
$$

\noindent If the quantity $\Omega^{\sigma[\nu \rho] }(x)$ is chosen as
$$
\Omega^{\sigma[\nu \rho] }(x)  =
+ {1 \over 2} \;
\bar{\Psi}  [ \Gamma^{\sigma}  J^{\nu \rho} +
J^{\nu \rho} \Gamma^{\sigma}  ] \; \Psi   \; ,
\eqno(9.29b)
$$

\noindent then  second and third terms on the right  in  (9.29a) cancel each other, and we reach
a conservation law
$$
\nabla_{\alpha} \; \bar{T}^{\;\;\; \alpha}_{\beta}(x) =
- i g \; J^{\alpha} (x) F_{\alpha \beta} (x) \;   .
\eqno(9.30)
$$

\subsection*{Results}

A generalized  vector particle theory with the use of an extended set of Lorentz group irredicible
representations, including scalar, two 4-vectors, and antisymmetric 2-rang tensor, is investigated
both in tensor an in matrix approaches. Initial equations depend upon four complex parameters $\lambda_{i}$,
obeying two sup\-plemen\-ta\-ry conditions, so restriction of the model to the case of electrically
neutral vector particle  is not a trivial task. A special basis in the space of 15-component wave functions
is found where instead of four $\lambda_{i}$ only one real-valued  quantity $\sigma$,
a bilinear combination of $\lambda_{i}$, is presented.  This $\lambda$-parameter is interpreted as an additional electromagnetic
characteristic of a charged vector particle, polarizability. It is shown that in this basis $C$-operation
is reduced to the complex conjugation only, without any accompanying linear transformation.
The form of $C$-operation in the initial
basis is calculated too. Invariant bilinear form matrix in both bases are found and the Lagrange formulation
of the whole theory is given. Explicit expressions of the conserved current vector  and
of the energy-momentum tensor are established. In presence of external electromagnetic fields, two
supple\-men\-ta\-ry field components, scalar and 4-vector, give a non-trivial contribution to the Lagrangian and
conserved quantities. Restriction to a  massless vector particle is determined.

Extension of the whole theory to the case of Riemannian space-time is accomp\-li\-shed. Two methods of obtaining
corresponding generally covariant wave equations are elaborated: of tensor- and of tetrad-based ones. Their equivalence is proved. It is shown that in case of pure curved space-time
models without Cartan torsion no specific additional interaction terms because of non-flat geometry arise.
The conformal symmetry of a massless generally covariant equation is demonstrated explicitly.
A canonical tensor of energy-momentum $T_{\beta \alpha}$  is constructed, its conservation law happens to involves
the Riemann curvature tensor. Within the framework of known ambiguity of any energy-momentum tensor,
a new tensor $\bar{T}_{\beta \alpha}$ is suggested to be  used, which obeys a common conservation law.

\newpage

\begin{center}
{\bf References }
\end{center}

\noindent 1.
Pauli W., Fierz M. Uber relativistische
Feldgleichungen von Teilchen mit beliebigem Spin im
electro\-magnetishen Feld. // Helv. Phys. Acta. 1939, Bd. 12, S.
297-300.

\noindent
2.  Fierz V., Pauli W. On relativistic wave equations
for particles of arbitrary spin in an electromagnetic  field. //
Proc. Roy. Soc. London. 1939, Vol. A173, P. 211-232.

\noindent
3.
Dirac P. Relativistic wave  equations. Proc. Roy. Soc. (London). A. 1936. Vol. 155. P. 447-459.

\noindent
4.
H.J. Bhabha. Relativistic wave equations for elementary particles.
Rev. Mod. Phys. 1945. Vol. 17. P. 200-216; 1949. vol. 21. P. 451;
On the postulational basis  of the theory  of elementary particles.  Rev. Mod. Phys. 1949. Vo;. 21.
P.  451-462.

\noindent
5.
Harish-Chandra. Phys. Rev. 1947. Vol. 71. P. 793; Proc. Roy. Soc. (London). 1947.
Vol. 192A. P. 195.

\noindent
6.
H. Umezawa. Quantum field theory. Amsterdam. 1956.

\noindent
7.
E.M. Corson. Introduction to tensors, spinors and relativistic wave equations. Blackie. London. 1953.

\noindent
8.
I.M.  Gel'fand, R.A. Minlos, and Z.Ya. Shapiro. Linear representations of the  rotation and Lorentz groups
and their applications. Pergamon. New York. 1963.

\noindent
9.
M.A. Naimark. Linear representations of the Lorentz group.  Pergamon. New York, 1964.

\noindent
10.
Krajcik   R., Nieti M.M.  Bhabha firdt-order wave equations. Phys. Rev. D. 1974. Vol.  10. P.
4049-4062; Phys. Rev. D. 1975. Vol.  11. P. 4042-1458;
Phys. Rev. D. 1975. Vol.  11. P. 4059-1471;
Phys. Rev. D. 1976. Vol.  13. P. 924-941;
Phys. Rev. D. 1976. Vol.  14. P. 418-436;
Phys. Rev. D. 1977. Vol.  15. P. 433-444;
Phys. Rev. D. 1977. Vol.  15. P. 445-452.

\noindent
11.
A.A. Bogush, L.G. Moroz.
{\it Introduction to theory of classical fields.}
Minsk,  1968.  385 pages (in Russian).

\noindent
12.
F.I. Fedorov. {\it The Lorentz group}.  Moskow,  1989.  384 pages (in Russian ).

\noindent
13.
V.L. Ginzburg.
{\it On theory of exited spin states of elementary particles.}
JETP. 1943. VOl. 13. P. 33-58 (in Russian).

\noindent
14.
E.S. Fradkin.
{\it On theory of particles with higher spins.}
JETP. 1950. Vol 20. P. 27-38 (in Russian).

\noindent
15.
Petras M.
{\it A note to Bhabha's equation for a particle with maximum spin 3/2.}
Czehc. J. Phys. 1955. Vol. 5. No 3. P. 418-419.

\noindent
16.
V.Ya. Fainberg.
{\it On interaction theory of higher spin particles  with electromagnetic  and meson fields.}
Trudy  FIAN SSSR. 1955. Vol. 6. P. 269-332 (in Russian).

\noindent
17.
I. Ulehla  {\it Anomalous equations for particles with spin 1/2.}
JETP. 1957. Vol. 33. P. 473-477 (in Russian).

\noindent
18.
Formanek J.
{\it On the Ulehla-Petras wave equation.}
Czehc. J. Phys. B.  1955. Vol. 25. No 8. P. 545-553.

\noindent
19.
S.I. Lobko.
{\it On theory particles with variable spin 1/2 -3/2 and with two rest masses.}
Kandid. Dissertation. Minsk,  1965 (in Russian).

\noindent
20.
F.I. Fedorov, V.A. Pletuschov.
{\it Wave equations with repeated representations of the Lorentz group. Integer spin.}
Vesti AN BSSR.  Ser. fiz.-mat. nauk.  1969. No 6. P. 81-88 (in Russian).

\noindent
21.
A.Z. Capri.
{\it Non-uniqueness of the spin 1/2 equation.} Phys. Rev. 1969.  Vol. 178. No 5. P. 1811-1815.

\noindent
22.
A.Z. Capri. {\it First-order  wave equations for half-odd-integral spin.}
Phys. Rev. 1969.  Vol. 178. No 5. P. 2427-2433.

\noindent
23.
A.Z. Capri. {\it First-order  wave equations for multi-mass fermions.}
Nuovo Cimento. B.  1969.  Vol. 64. No 1. P. 151-158.

\noindent
24.
F.I. Fedorov, V.A. Pletuschov. {\it
Wave equations with repeated representations of the Lorentz group. Half-integer spin.}
Vesti AN BSSR. Ser. fiz.-mat. nauk.  1970. No 3. P. 78-83 (in Russian).

\noindent
25.
V.A. Pletuschov, F.I. Fedorov.
{\it Wave equations with repeated representations of the Lorentz group for a spin 0 particle.}
Vesti AN BSSR. Ser. fiz.-mat. nauk.  1970.  No 2. P. 79-85 (in Russian).

\noindent
26.
V.A. Pletuschov, F.I. Fedorov.
{\it Wave equations with repeated representations of the Lorentz group for a spin 1 particle.}
Vesti AN BSSR. Ser. fiz.-mat. nauk. 1970,  No 3. P. 84-92 (in Russian).

\noindent
27.
A. Shamaly, A.Z. Capri. {\it First-order wave equations for integral spin.}
Nuovo Cimento. B. 1971. Vol. 2. No 2. P. 235-253.

\noindent
28.
A.Z. Capri.
{\it Electromagnetic properties of a new spin-1/2 field.}
Progr. Theor. Phys.  1972. Vol. 48. No 4. P. 1364-1374.

\noindent
29.
A. Shamaly, A.Z. Capri.
{\it Unified theories for massive spin 1 fields.}
Can. J, Phys. 1973. Vol. 51. No 14. P. 1467-1470.

\noindent
30.
M.A.K. Khalil.
{\it  Properties of a 20-component spin 1/2 relativistic wave equation. }
Phys. Rev. D.  1977.  Vol. 15. No 6. P. 1532-1539.

\noindent
31.
M.A.K. Khalil.
{\it Barnacle equivalence structure in relativistic wave equation.}
Progr. Theor. Phys.  1978. Vol. 60. No 5. P. 1559-1579.

\noindent
32.
M.A.K. Khalil.
{\it An equivalence of relativistic field equations.}
Nuovo Cimento. A. 1978. Vol. 45. No  3. P. 389-404.

\noindent
33.
M.A.K. Khalil.
{\it Reducible  relativistic wave equations.}
J. Phys. A.: Math. and Gen.  1979. Vol. 12. No  5. P. 649-663.

\noindent
34.
A.A. Bogush, V.V. Kisel.
{\it Equations with repeated representations of the Lorentz group and Pauli interaction.}
Vesti AN BSSR. Ser. fiz.-mat. nauk.   1979.  No 3.  P. 61-65 (in Russian).

\noindent
35.
A.A. Bogush, V.V. Kisel, M.I. Levchuk, L.G. Moroz.
{\it On description of polarizability of scalar particles in the theory of relativistic wave equations.}
In: Covariant methods in  theoretical physics. Elementary particle physics and relativity theory.
Institute of Physics, Academy of sciences of Belarus. Minsk. 1981.  P. 81-90 (in Russian).

\noindent
36.
V.V. Kisel.
{\it Electric polarizability of a spin 1 particle in the theory of relativistic wave  equations.}
Vesti AN BSSR. Ser. fiz.-mat. nauk.  1982. No  3. P. 73-78 (in Russian).

\noindent
37.
A.A. Bogush, V.V. Kisel.
{\it Description of a free particle by different wave equations.}
Doklady AN BSSR.  1984. Vol. 28. No 8. P. 702-705 (in Russian).

\noindent
38.
A.A. Bogush, V.V. Kisel.
{\it Equation for a spin 1/2 particle with anomalous magnetic momentum.}
// Izvestiya VUZov. Fizika. 1984.  No 1. P. 23-27 (in Russian).

\noindent
39.
A.A. Bogush, V.V. Kisel, F.I. Fedorov.
{\it On interpretation of supplementary components of wave functions in presence of electromagnetic
interaction.}
Doklady AN BSSR.  1984. Vol. 277. No 2.  P. 343-346 (in Russian).

\noindent
40.
Lunardi J.T., Pimentel B.M.,  Teixeira R.G.
{\em Duffin-Kemmer-Petiau equation in Rieman\-nian space-times.} -- 14 pages. gr-qc/9909033.

\noindent
41.
Lunardi J. T., Pimentel B.M., Teixeira R. G. and Valverde J.S. 2000.
{\em Remarks on Duffin-Kemmer-Petiau theory and guage invariance.} Phys. Lett. A. 268. 165-73.

\noindent
42.
Fainberg V. Ya. and Pimentel B.M.  {\em Duffin-Kemmer-Petiau  and Klein-Gordon-Fock
equations for electromagnetic, Yang-Mills and external gravitational field interactions:
proof of equivalence.} Phys. Lett. 2000.  A. Vol. 271. P. 16-25.

\noindent
43.
Fainberg V. Ya. and Pimentel B.M. {\em On equivalence Duffin-Kemmer-Petiau
and Klein-Gordon equations.}  Braz. J. Phys. 2000. Vol. 30.  P. 275-81.

\noindent
44.
Fainberg V. Ya. and Pimentel B.M.
{\em Equivalence Duffin-Kemmer-Petiau and Klein-Gordon-Fock equations.} Theor. Math. Phys. 2000.
Vol. 124. P. 1234-49.

\noindent
45.
de Montigny M., Khanna F.C., Santana A. E., Santos E.S. and Vianna J.D.M.
{\em Galilean covariance and the Duffin-Kemmer-Petiau equation.}  J.Phys. A.: Math.
Gen. 2000. Vol. 33. L273-8.

\noindent
46.
Tetrode H. {\em Allgemein relativistishe Quantern theorie des Elektrons.}  Z.
Phys. 1928, Bd 50, S. 336.

\noindent
47.
 Weyl H.
{\em Gravitation and the electron.}
Proc. Nat. Acad. Sci. Amer. -- 1929. -- Vol. 15. -- P. 323 -334;
{\em Gravitation and the electron.}
Rice Inst. Pamphlet.  -- 1929. --  Vol. 16. --  P. 280-295;
{\em Elektron und Gravitation.}  Z. Phys. -- 1929. --  Bd. 56. --  S. 330-352;
{\em A remark on the coupling of gravitation and electron.}
 Actas   Akad. Nat. Ciencias  Exactas. Fis. y natur. Lima. -- 1948. --
Vol. 11. -- P. 1-17.

\noindent
48.
 Fock V., Ivanenko D.
{\em \"{U}ber eine m\"ogliche geometrische Deutung der relativistischen
Quanten\-theorie.}  Z. Phys.  -- 1929. -- Bd. 54. -- S. 798-802;
{\em G\'{e}ometrie   quantique  lin\'{e}aire   et d\'{e}placement parallele.}
 Compt. Rend. Acad. Sci. Paris. --  1929. -- Vol. 188. -- P. 1470-1472.

\noindent
49.
 Fock V.
{\em Geometrisierung der Diracschen Theorie des Elektrons.}
Z. Phys. -- 1929. -- Bd. 57, n. 3-4. -- S. 261-277;
{\em Sur les \'{e}quations de Dirac dans la th\'{e}orie de relativit\'{e}
g\'{e}n\'{e}rale.}
 Compt. Rend. Acad. Sci. Paris. -- 1929. -- Vol. 189. -- P. 25-28.

\noindent
50.
Schouten J.A.
{\em Dirac  equations  in  general  relativity.}
J. Math. and Phys. -- 1931. -- Vol. 10. --  P. 239-271; P. 272-283.

\noindent
51.
 Schr\"odinger E.
{\em Sur la th\'{e}rie relativiste de l'\'{e}lectron et l'interpr\'{e}tation
de la m\'{e}chanique  quantique.}
Ann. Inst. H. Poincar\'{e}. -- 1932. -- Vol. 2. -- P. 269-310;
{\em Dirac'sches Elektron  im  Schwerfeld.}
Sitz. Ber. Preuss. Akad. Wiss. Berlin. Phys.-Math Kl.
 -- 1932. --  S. 105-128.

\noindent
52.
 Einstein  A.,  Mayer  W.
{\em Semivektoren   und   Spinoren.}
Sitz. Ber. Preuss. Akad. Wiss. Berlin. Phys.-Math. Kl.
-- 1932 . -- S. 522-550;
{\em Die Diracgleichungen f\"ur Semivektoren.}
Proc. Akad.  Wet. (Amsterdam). -- 1933. -- Bd. 36. -- S. 497-516;
{\em Spaltung der Nat\"urlichsten  Feldgleichungen  f\"ur
Semi-Vektoren  in  Spinor-Gleichungen  von  Diracschen Tipus.}
Proc. Akad. Wet. (Amsterdam). -- 1933. -- Bd. 36. -- S. 615-619.

\noindent
53.
 Bargmann V.
{\em \"{U}ber der Zusammenhang zwischen  Semivektoren  und
Spinoren  und  die   Reduktion   der   Diracgleichungen   fur Semivektoren.}
Helv. Phys. Acta. -- 1933. -- Bd. 7. -- S. 57.

\noindent
54.
 Schouten J.A.
{\em Zur generellen Feldtheorie,  Semi-Vektoren  und Spin-raum.}
Z. Phys. -- 1933. -- Bd. 84. -- S. 92-111.

\noindent
55.
Infeld L., van der Waerden B.L.
Die  Wellengleichungen  des
{\em Elektrons in der allgemeinen Relativit\"{a}stheorie.}
Sitz. Ber. Preuss. Akad. Wiss. Berlin. Phys.-Math. Kl. -- 1933. --
Bd. 9. -- S. 380-401.

\noindent
56.
 Infeld L.
{\em Dirac's  equation  in  general  relativity  theory.}
Acta Phys. Polon. -- 1934. -- Vol. 3, n. 3. --  P. 1.

\noindent
57.
 Dirac P.A.M.
{em Wave  equations in conformal space.}
Ann. Math. -- 1936. -- Vol. 37. -- P. 429-442.

\noindent
58.
 Yamamoto H.
{\em On equations for the Dirac  electron in  general relativity.}
Japan. J. Phys. --  1936. -- Vol. 11 .--  P. 35.

\noindent
59.
 Proca A.
{\em Sur la th\'{e}orie ondulatoire des electrons positifs et n\'{e}gatifs.}
J. Phys. et Rad. -- 1936. -- Vol. 7. -- P. 347.

\noindent
60.
 Benedictus W.
{\em Les \'{e}quations de Dirac dans un  espace \`{a} m\'{e}trique
riemannienne. }
Compt. Rend. Acad. Sci. Paris. -- 1938.  -- Vol. 206. -- P. 1951.

\noindent
61.
 Cartan E.
{\em Le\c{c}ons sur la th\'eorie des  spineurs.}
 Actualit\'es Sci. -- 1938.

\noindent
62.
 Pauli W.
{\em \"Uber die Invarianz  der Dirac'schen Wellengleichungen gegen\"uber
\"Ahnlichkeit\- stransformationen des Linienelementes im Fall verschwindender
Ruhmasse.}
Helv. Phys. Acta. -- 1940. -- Bd. 13. -- S. 204-208.

\noindent
63.
 Bade W.L., Jehle H.
{\em An introduction to spinors.}
Rev. Mod. Phys. -- 1953. -- Vol. 25. -- P. 714-728.

\noindent
64.
Brill D.R.,    Wheeler J.A.
{\em Interaction of neutrinos   and gravitational fields.}
Rev. Mod. Phys. -- 1957. -- Vol. 29,  n. 3. --  P. 465-479.

\noindent
65.
Bergmann  P.G.
{\em Two-component   spinors   in   general relativity.}
Phys. Rev. -- 1957. -- Vol. 107, n. 2. -- P. 624-629.

\noindent
66.
 Fletcher J.G.
{\em Dirac matrices in Rimannian space.}
Nuovo Cim. --  1958. --  Vol. 8, n. 3. -- P. 451-458.

\noindent
67.
 Buchdahl H.A.
{\em On extended  conformal transformations of spinors and spinor
equations.}
Nuovo Cim.  -- 1959. -- Vol. 11. -- P. 496-506.

\noindent
68.
 Namyslowski J.
{\em The Dirac equation in general relativity in the vierbein formalism.}
Acta Phys. Polon. -- 1961. -- Vol. 20, n. 11. -- P. 927-936;
{\em Symmetrical form of Dirac matrices in general relativity.}
Acta Phys. Polon. -- 1963. -- Vol. 23, n. 6. -- P. 673-684.

\noindent
69.
Peres A.
{\em Spinor fields in generally covariant theories.}
Nuovo Cim. Suppl. -- 1962. -- Vol. 24, n. 2. -- P. 389-452.

\noindent
70.
Lichnerowicz  A.
{\em Champ  de  Dirac,  champ du  neutrino  et
transformations $C,P,T$ sur un espace-temps curve.}
Ann. Inst. Henri Poincar\'e. A.  1964, Vol. 1, n. 3,  P. 233-290.

\noindent
71.
Ogivetckiy V.I., Polubarinov I.V. {\em On spinors in gravity theory} (in Russian).
JETP.  1965. Vol 48, n. 6. P. 1625-1636.

\noindent
72.
Pagels H.
{\em Spin and Gravitation.}
Ann. Phys. (N.Y.)-- 1965. -- Vol. 31, n. 1. --  P. 64-87.

\noindent
73.
Brill D.R., Cohen J.M.
{\em Cartan frames and general relativistic Dirac equation.}
J. Math. Phys. -- 1966. -- Vol. 7,  n. 2. -- P. 238-245.

\noindent
74.
Cap F., Majerotto W., Raab W., Unteregger P.
{\em Spinor  calculus in Riemannian manifolds.}
Fortschr. der  Physik.  -- 1966. -- Bd. 14, n. 3. -- S. 205-233.

\noindent
75.
Von D. Kramer,  Stephani H.
{\em Bispinorfelder im Riemannschen  Raum.}
Acta Phys. Polonica. -- 1966. -- Vol. 29, n. 3. -- P. 379-386.

\noindent
76.
Luehr C.P., Rosenbaum M.
{\em Spinor connections in general relativity.}
J. Math. Phys. -- 1974. -- Vol. 15, n. 7. -- P. 1120-1137.

\noindent
77.
Maia M.D.
{\em Conformal  spinor fields in general relativity.|
J. Math. Phys. -- 1974. -- Vol. 15, n. 4. -- P. 420-425.

\noindent
78.
Brum\^a C. {|em A mode of constructing the  Dirac matrices in gravitational
field.}  Rev. Roum. Phys. -- 1986. -- Tome 31,  n. 8. -- P. 753-763;
{\em  On Dirac matrices in gravitational field.}
 Rev. Roum. Phys. -- 1987. -- Tome 32,  n. 4. -- P. 375-382.

\noindent
79.
Fariborz A.H., McKeon D.G.C. {\em  Spinors in Weyl geometry.}
Class. Quant. Grav. -- 1997. -- Vol. 14. -- P. 2517-2525. -- hep-th/9607056.

\noindent
80.
Manoelito  M. de Souza. {\em  The Lorentz Dirac equation and the structure of
spacetime.}   Braz. J. Phys. -- 1998. -- Vol. 28. -- P. 250-256. --
 hep-th/9505169.

\end{document}